\documentclass[draftcls, onecolumn, peerreview, journal]{IEEEtran}
\usepackage{amssymb}
\usepackage{amsfonts}
\usepackage{amsbsy}
\usepackage{amsmath}
\usepackage{amssymb}
\usepackage{amsthm}
\usepackage{algorithmic}
\usepackage{algorithm}
\usepackage{graphicx}
\usepackage{subfig}
\usepackage{rotating}

\newtheorem{thm}{Theorem}
\newtheorem{lem}[thm]{Lemma}
\newtheorem{example}[thm]{Example}
\newcommand{\pf}{\paragraph*{Proof}}

\newcommand{\pfend}{\par\vspace{2ex}\noindent}
\newcommand{\eind}{\hspace*{\fill}$\Box$\par\vspace{2ex}\noindent}

\newcommand{\ruimte}{\par\vspace{1.0ex}}

\DeclareMathOperator*{\argmin}{argmin}

\title{\LARGE \bf Modified Rice-Golomb Code for Predictive Coding of Integers with Real-valued Predictions}

\author{Mortuza Ali and Manzur Murshed,~\IEEEmembership{Member,~IEEE}% <-this % stops a space
\thanks{M. Ali and M. Murshed are with Gippsland School of Information Technology, Monash University, Victoria, 3842, Australia.

e-mails: (mortuza.ali@monash.edu; manzur.murshed@monash.edu)

This research is supported by the Australian Research Council (ARC).
}% <-this % stops a space
}

\begin{document}

\maketitle

%\IEEEpeerreviewmaketitle

%%%%%%%%%%%%%%%%%%%%%%%%%%%%%%%%%%%%%%%%%%%%%%%%%%%%%%%%%%%%%%%%%%%%%%%%%%%%%%%%
\begin{abstract}
Rice-Golomb codes are widely used in practice to encode integer-valued prediction residuals. However, in lossless coding of audio, image, and video, specially those involving linear predictors, the predictions are from the real domain. In this paper, we have modified and extended the Rice-Golomb code so that it can operate at fractional precision to efficiently exploit the real-valued predictions. Coding at arbitrarily small precision allows the residuals to be modeled with the Laplace distribution instead of its discrete counterpart, namely the two-sided geometric distribution (TSGD). Unlike the Rice-Golomb code, which maps equally probable opposite-signed residuals to different integers,  the proposed coding scheme is symmetric in the sense that, at arbitrarily small precision, it assigns codewords of equal length to equally probable residual intervals. The symmetry of both the Laplace distribution and the code facilitates the analysis of the proposed coding scheme to determine the average code-length and the optimal value of the associated coding parameter. Experimental results demonstrate that the proposed scheme, by making efficient use of real-valued predictions, achieves better compression as compared to the conventional scheme.
\end{abstract}

\section{Introduction}
Prediction, which involves estimating the outcome of a data source given some past observations,
%~\cite{GershoGrayVQ}
 is an effective tool for data compression. Consider the sequential encoding of an integer-valued source $\{ X_t\}$, $t=0, 1, \ldots$, over the alphabet $\{0, 1, \ldots, (q-1)\}$, on a symbol-by-symbol basis. In predictive coding, given the previously encoded data $x_0, x_1, \ldots, x_{t-1}$, the value of $x_t$ is predicted as $\hat{x}_t$ and the residual $\varepsilon_t = x_t - \hat{x}_t$ is encoded using an entropy code such as Huffman code or Rice-Golomb code. Since the previously encoded data are already available at the decoder, it can make the same prediction $\hat{x}_t$ and reconstruct $x_t = \hat{x}_t + \varepsilon_t$. The beneficial effect of prediction is that it decorrelates the data in the sense that the entropy of the residuals is significantly lower than the entropy of the original sequence.
\ruimte

Although nonlinear predictions place no restriction on the predictor and thus can achieve better performance, linear predictions that restrict the predictor to be a linear function of the previously encoded data are much simpler to construct and analyze. Therefore, linear predictors are widely used in practice. In $\alpha$ order linear predictive coding, after having observed the past data sequence $x_0, x_1, \ldots, x_{t-1}$, the value of $x_t$ is predicted as a linear combination of the previous $\alpha$ values as
\begin{equation}
\hat{x}_t = \sum_{j=0}^{\alpha-1}a_j x_{t-j}
\end{equation}
where $a_j \in \mathbb{R}, 0 \leq j < \alpha$, are real-valued predictor coefficients that need to be optimized. The most common measure of the performance of a predictor is the {\em mean squared error (MSE)}. Therefore, the coefficients $a^*_j$'s that minimize the MSE, i.e., \mbox{$E[\|X- \hat{X}\|^2]$}, are considered optimal. In practice, however, $a^*_j$'s are learnt by minimizing the MSE over a window of size $\omega$ such that
\begin{equation}
(a^*_0, a^*_1, \ldots, a^*_{\alpha-1}) = \argmin_{(a_0, a_1, \ldots, a_{\alpha-1}) \in \mathbb{R}^\alpha} \sum_{i=1}^{\omega} \left ( x_{t-i} - \sum_{j=0}^{\alpha -1}a_j x_{t-i-j} \right )^2.
\end{equation}
Note that even if the source values $x_i$'s are integers, the predictor coefficients $a^*_j$'s are drawn from the real domain $\mathbb{R}$ leading to real-valued predictions and residuals.
\ruimte

Conventional predictive coding techniques round the real-valued prediction $\hat{x}$\footnote{The subscript $t$ has been dropped for notational convenience.} to the nearest integer $[\hat{x}]$ and then encode the integer residual
\begin{equation}
\bar{\varepsilon} = x -[\hat{x}].
\end{equation}
Given that $[\hat{x}]$ is also available at the decoder, we can restrict the number of possible values for $\bar{\varepsilon}$ to $q$ by taking into account the fact that $x$ can only take values from the alphabet $\{0, 1, \ldots, (q-1) \}$~(see \cite{LOCOI:Jour}). Now encoding of $\bar{\varepsilon}$  using an entropy coder requires the knowledge of the probability distribution of the residuals. Therefore, in sequential symbol by symbol coding, the probabilities of the $q$ possible residual values are also needed to be estimated adaptively. However, for a large alphabet there might not be sufficient number of data in practice to reliably estimate these probabilities. Hence, in practical applications, a parametric representation of the probability distribution of the residuals is often preferred~\cite{LOCOI:Jour, CommLett:OptPrefix}.
\ruimte

It has been observed that the distributions of the real-valued prediction residuals in audio~\cite{RobinsonSHORTEN}, image~\cite{HowardVitter}, and video~\cite{CommLett:NealTV,VideoEntropyCriterion} coding highly peak at zero, that can be closely approximated by the Laplace distributions. A Laplace distribution, which sharply peaks at zero, is defined by the following probability density function (pdf),
\begin{equation}
\label{eqn:LapPdf}
f_{\theta}(\varepsilon) = -\frac{\ln\theta}{2}\theta^{|\varepsilon|}, \quad 0<\theta<1, \quad \varepsilon \in \mathbb{R}.
\end{equation}
Here $\theta$ is a scale parameter which controls the two-sided decay rate. Since conventional predictive coding schemes encode the integer-valued residuals, they model the distribution of $\bar{\varepsilon}$ using a`discrete analog' of the Laplace distribution. A discrete analog of a continuous distribution with pdf $f_{\theta}(\cdot)$ has been proposed in~\cite{CommLett:DisAna} as a discrete distribution supported on the set $\mathbb{Z}$  having the probability mass function (pmf)
\begin{equation}
\label{eqn:discreteAna}
p_{\theta}(i) = \frac {f_{\theta}(i)}{\sum_{j = - \infty}^{\infty}f_{\theta}(j)}, \quad i, \ j \in \mathbb{Z}.
\end{equation}
It follows from~(\ref{eqn:LapPdf}) and~(\ref{eqn:discreteAna}) that the pmf of the discrete analog of the Laplace distribution takes the form,
\begin{equation}
p_{\theta }(i) = \frac {1 - \theta}{1 + \theta} \theta^{|i|}, \quad i \in \mathbb{Z}.
\end{equation}
which is known as the two-sided geometric distribution (TSGD) in the literature~\cite{CommLett:OptPrefix}.
\ruimte

 Popular prefix coding schemes use Golomb codes~\cite{GolombRunLen} to exploit the exponential decaying in the pmf of integer residuals. However, Golomb codes are optimal~\cite{CommLett:GeomOpt} for the one-sided geometric distribution (OSGD) of the form
\begin{equation}
p_{\theta }(i) = (1 - \theta)\theta^{i}, \quad i \geq 0.
\end{equation}
Given a positive integer parameter $m_g$, the Golomb code of $i$ has two parts: the prefix $\lfloor i/m_g \rfloor$ in unary representation and the reminder of that division, $i \textnormal{ mod } m_g$, in minimal binary representation. The unary representation of a non-negative integer $j$ consists of $j$ number of `1's followed by a `0'. The minimal binary representation of a non-negative integer $k$ from the alphabet $\{0, 1, \ldots, m_g-1\}$ uses $\lfloor \lg m_g \rfloor$ bits when $k < 2^{\lceil \lg m_g \rceil} - m_g$ or $\lceil \lg m_g \rceil$ bits otherwise. For a given $\theta$, the optimal value of the parameter $m_g$ is given by~\cite{CommLett:GeomOpt}
\begin{equation}
m_g^* = \left \lceil \frac {\lg (1 + \theta)} {- \lg \theta} \right \rceil.
\end{equation}
Since Golomb codes are defined for non-negative integers only, popular Golomb-based codecs map the integer residual $\bar{\varepsilon}$ into a unique non-negative integer prior to encoding by the following overlap and interleave scheme originally proposed by Rice in~\cite{RiceMapping},
\begin{equation}
M_{\textnormal{Rice}}(\bar{\varepsilon}) = \left \{ \begin{array}{ll} 2\bar{\varepsilon}, & \bar{\varepsilon} \geq 0; \\ -2\bar{\varepsilon}-1, & \textrm{otherwise.} \end{array} \right.
\label{eqn:riceMapping}
\end{equation}
Although Golomb codes are optimal for geometrically distributed non-negative integers, the above mentioned scheme with Rice-mapping is not optimal for TSGD (see~\cite{CommLett:OptPrefix}). Moreover, when the predictions are constrained to integers, a real-valued bias is typically present in the prediction residuals, which introduces a shift parameter $\mu$ in the TSGD model in addition to the parameter $\theta$ (see~\cite{LOCOI:Jour} and~\cite{WuMemonCALIC}). This off-centred TSGD is modeled by the following pmf~\cite{CommLett:OptPrefix}
\begin{equation}
p_{\theta, \mu}(i) = \frac {1 - \theta}{\theta^{1 - \mu} + \theta^{\mu}} \theta^{|i + \mu|}, \quad 0 \leq \mu < 1, \quad i \in \mathbb{Z}.
\end{equation}

A complete characterization of the optimal prefix code for the off-centred TSGD has been presented in~\cite{CommLett:OptPrefix}. It divides the two-dimensional parameter space $(\theta, \mu)$ into four types of region and associates a different code construct with each type. However, the article admits that two dimensionality of the parameter space adds significant complexity to the characterization and analysis of the code. Moreover, being optimal for the off-centred TSGD, these codes do not preclude improving predictive compression gain further if the residuals could be handled in the real domain, minimizing the loss due to rounding.
\ruimte

In this paper, we extend and modify the Rice-Golomb code so that it can handle the real-valued predictions at an arbitrary precision. More specifically, the contribution of the paper can be summarized as follows. Firstly, we generalize the Rice mapping~(\ref{eqn:riceMapping}) so that it can operate at an arbitrary precision. We then present the complete encoding and decoding algorithms based on the generalized Rice mapping. With the generalized Rice mapping, although the encoding is similar to Rice-Golomb encoding, the decoding is slightly convoluted. One of the salient features of the proposed coding scheme is that it is symmetric, i.e, when operating at finest precision, it assigns codewords of equal length to equally probable residual intervals. Secondly, assuming that the real-valued residuals are Laplace distributed, we analyze the proposed coding scheme and determine the close form expression for the average code length. Thirdly, we determine the relationship between the scale parameter $\theta$ and the optimal value of the coding parameter $m$ when the code operates at the finest precision. The analysis of the proposed scheme to determine the average code-length and the optimal value of the associated coding parameter is greatly facilitated by the symmetry in both the Laplace distribution and the code. Although, for the codes operating at other precision, a relationship between $\theta$ and $m$ is not readily available, we have demonstrated analytically and experimentally that a sub-optimal strategy incurs negligible redundancy at sufficiently small precision.

\ruimte
The organization of the rest of the paper is as follows. The modified Rice-Golomb code is presented along with a novel generalized Rice mapping and the implementation details of the proposed coding scheme in Section~\ref{formulation}. In Section~\ref{MRGanalysis}, the proposed scheme is then analyzed to determine its average code-length and the relationship between the scale parameter $\theta$ and the coding parameter $m$ at different fractional precisions. Finally, experimental results are presented to demonstrate the efficacy of the proposed coding scheme in Section~\ref{sec:exp}.

\section{Modified Rice-Golomb code at fractional precision}
\label{formulation}
In conventional predictive coding of integers, the real-valued predictions are first mapped to their nearest integers and then the integer-valued residuals are encoded with an entropy code. In order to extend the Rice-Golomb code to exploit the real-valued prediction at any arbitrary precision, let the scheme operate at precision $\rho/\tau$, where $\rho$ and $\tau$ are positive integers and $\rho \leq \tau$. Therefore, prior to residual encoding, the prediction $\hat{x}$ is rounded to $[\hat{x}]_{\rho/\tau} = \rho\lfloor \tau\hat{x}/\rho + 1/2 \rfloor /\tau$, which is the integer multiple of $\rho/\tau $ nearest to $\hat{x}$. Standard Rice-Golomb code is then an instance of this extended code with $\rho=\tau=1$.
\ruimte

Although $[\hat{x}]_{\rho/\tau }$ can take any integer multiple value of $\rho/\tau $, using the fact that unknown $x$ is from $\mathbb{Z}$, the decoder can deduce that the residual $\bar{\varepsilon}_{\rho/\tau } = x - [\hat{x}]_{\rho/\tau }$ will take integer-apart values in the form
\begin{equation}
\label{eqn:gamma_frac}
\varepsilon_\gamma = \gamma + \Delta/\tau \quad \textnormal{where} \, \gamma \in \mathbb{Z} \,\, \textnormal{and} \,\, \Delta \, \textnormal{is a fixed integer from} \quad [0, \tau).
\end{equation}
\begin{example} When $\rho=1$ and $\tau = 4$, the possible residual values are integer multiples of $\rho/\tau =1/4$, i.e., $\{ 0, \pm 1/4, \pm 2/4, \pm 3/4, \ldots \}$. Let the prediction be $\hat{x}=0.70$, which must be rounded to the nearest multiple of $1/4$, i.e., to $[\hat{x}]_{1/4} = \lfloor 4 \times 0.70 + 1/2 \rfloor / 4 = 0.75$. Once the rounded prediction $[\hat{x}]_{1/4}$ is fixed to $0.75$, the possible residual values are $\{ \ldots, -1.75, -0.75, 0.25, 1.25, \ldots\}$, which are of the form $\varepsilon_\gamma = \gamma + 1/4, \gamma \in \mathbb{Z}$.
\end{example}
\ruimte

Now these integer-apart discrete residual values need to be mapped to unique non-negative integers so that they can be encoded using Golomb codes. For an efficient implementation of the code, the mapping also needs to be easy to compute.

\subsection{Residual mapping}
According to the Laplace distribution, small-valued residuals have higher probabilities than those of large-valued residuals. Since Golomb codes assign shorter-length codewords to small-valued non-negative integers, small-valued discrete residuals should be mapped to small-valued non-negative integers. In the above example, $0.25$ should be mapped to $0$, and $-0.75$ should me mapped to $1$ and so on. More generally, if $\Delta < \tau/2$ then $|\varepsilon_0| < |\varepsilon_{-1}| < |\varepsilon_{1}| < |\varepsilon_{-2}| < \cdots$. Thus, $\varepsilon_\gamma$ should be mapped according to the function
\begin{equation}
M_{\Delta < \tau/2}(\varepsilon_\gamma) = \left \{ \begin{array}{ll} 2\gamma, & \gamma \geq 0; \\ -2\gamma -1, & \textrm{otherwise.} \end{array} \right.
\label{CommLett:Map_r_less}
\end{equation}
Similarly, if $\Delta \geq \tau/2$ then $|\varepsilon_{-1}| \leq |\varepsilon_{0}| \leq |\varepsilon_{-2}| \leq |\varepsilon_{1}| \leq \cdots$ and consequently, $\varepsilon_\gamma$ should be mapped according to the function,
\begin{equation}
M_{\Delta \geq \tau/2}(\varepsilon_\gamma) = \left \{ \begin{array}{ll} 2\gamma + 1, & \gamma \geq 0; \\ -2\gamma -2, & \textrm{otherwise.} \end{array} \right.
\label{CommLett:Map_r_greaterequal}
\end{equation}
The mapping~(\ref{CommLett:Map_r_less})  and (\ref{CommLett:Map_r_greaterequal}) require an explicit formula for the computation of $\gamma$ associated with the residual $\bar{\varepsilon}_{\rho/\tau } = x - [\hat{x}]_{\rho/\tau }$. It can be shown that
\begin{equation}
\bar{\varepsilon}_{\rho/\tau} = \lfloor\bar{\varepsilon}_{\rho/\tau}\rfloor + \lfloor\tau(\bar{\varepsilon}_{\rho/\tau} - \lfloor\bar{\varepsilon}_{\rho/\tau}\rfloor)\rfloor/\tau.
\label{CommLett:error_with_gamma}
\end{equation}
Obviously, $\lfloor\bar{\varepsilon}_{\rho/\tau}\rfloor \in \mathbb{Z}$ and  $\lfloor\tau(\bar{\varepsilon}_{\rho/\tau} - \lfloor\bar{\varepsilon}_{\rho/\tau}\rfloor)\rfloor < \tau$. Therefore, it follows from (\ref{eqn:gamma_frac}) and (\ref{CommLett:error_with_gamma}) that
\begin{equation}
\gamma = \lfloor\bar{\varepsilon}_{\rho/\tau}\rfloor
\end{equation}
and
\begin{equation}
\Delta = \lfloor\tau(\bar{\varepsilon}_{\rho/\tau} - \lfloor\bar{\varepsilon}_{\rho/\tau}\rfloor)\rfloor.
\end{equation}
When these values of $\gamma$ and $\Delta$ are substituted in (\ref{CommLett:Map_r_less}) and (\ref{CommLett:Map_r_greaterequal}), both the mappings converge to the following,
\begin{equation}
M(\bar{\varepsilon}_{\rho/\tau}) = \left \{ \begin{array}{ll} \lfloor 2\bar{\varepsilon}_{\rho/\tau} \rfloor, & \textnormal{if} \, \, \bar{\varepsilon}_{\rho/\tau} \geq 0; \\ -\lfloor 2\bar{\varepsilon}_{\rho/\tau}  \rfloor -1, & \textrm{otherwise.} \end{array} \right.
\label{CommLett:Map_unique}
\end{equation}
Indeed this mapping is the generalization of the Rice mapping~(\ref{eqn:riceMapping}). When $\rho = \tau=1$, the residual $\bar{\varepsilon}_{\rho/\tau} = \bar{\varepsilon}$ is integer valued and thus $\lfloor 2\bar{\varepsilon} \rfloor = 2\bar{\varepsilon}$. In this case, the mapping~(\ref{CommLett:Map_unique}) transforms into the Rice mapping~(\ref{eqn:riceMapping}). In the other extreme case of $\rho/\tau \to 0$, the prediction $\hat{x}$ is not rounded at all. For this asymptotic case of $\rho/\tau \to 0$, we will denote the mapping with $M(\varepsilon)$.

\subsection{Encoding and Decoding}
Having defined the mapping at precision $\rho/\tau$, the encoding operation is similar to the Rice-Golomb coding, however, the decoding operation is slightly convoluted due to the use of the floor function in the residual mapping~(\ref{CommLett:Map_unique}).
\ruimte
{\em Encoding: } Given the parameter value $m$, compute $j$ and $k$ as follows
\begin{eqnarray}
j = \lfloor M(\bar{\varepsilon}_{\rho/\tau})/m \rfloor,\\
k = M(\bar{\varepsilon}_{\rho/\tau}) \textnormal{ mod } m.
\end{eqnarray}
Then encode $j$ in unary and $k$ in minimal binary. The pseudocode of the encoding algorithm is given in Algorithm~\ref{Algo-MRGEncode}.
\begin{algorithm}[!htb]
\caption{Encoding}
\label{Algo-MRGEncode}
\begin{algorithmic}
\STATE
\STATE {\bf Input}: $x \in \mathbb{Z}, \hat{x} \in \mathbb{R}, \tau, \rho, m  \in \mathbb{Z}^+$
\STATE {\bf Output}: $j, k$.
%\STATE
\STATE $[\hat{x}]_{\rho/\tau} := \rho\lfloor \tau\hat{x}/\rho + 1/2 \rfloor /\tau$;
\STATE $\bar{\varepsilon}_{\rho/\tau } := x - [\hat{x}]_{\rho/\tau }$;
\IF{$\bar{\varepsilon}_{\rho/\tau } \geq 0$}
\STATE $M(\bar{\varepsilon}_{\rho/\tau})  :=  \lfloor 2\bar{\varepsilon}_{\rho/\tau} \rfloor$;
\ELSE
\STATE $M(\bar{\varepsilon}_{\rho/\tau})  :=  - \lfloor 2\bar{\varepsilon}_{\rho/\tau} \rfloor - 1$;
\ENDIF
\STATE $j := \lfloor M(\bar{\varepsilon}_{\rho/\tau})/m \rfloor$;\\
\STATE $k := M(\bar{\varepsilon}_{\rho/\tau}) \textnormal{ mod } m$;\\
\STATE Encode $j$ in unary and $k$ in minimal binary;\\
\end{algorithmic}
\end{algorithm}

\ruimte
{\em Decoding: } Given $m$, the decoder can compute $M(\bar{\varepsilon}_{\rho/\tau}) = jm + k$. Given $\hat{x}$, recovering $x$ from $M(\bar{\varepsilon}_{\rho/\tau})$, however, is not straightforward. In Rice-Golomb coding, by checking whether $M_{\textnormal{Rice}}(\bar{\varepsilon})$ is even or odd the decoder can decide which of the constituent functions in~(\ref{eqn:riceMapping}) was used: if $M_{\textnormal{Rice}}(\bar{\varepsilon})$ is even, it is the first constituent function that was used and $\bar{\varepsilon} = M_{\textnormal{Rice}}(\bar{\varepsilon})/2$; otherwise the second constituent function was used and $\bar{\varepsilon} = -(M_{\textnormal{Rice}}(\bar{\varepsilon})+1)/2$. Unlike the Rice mapping~(\ref{eqn:riceMapping}), where the value of the first constituent function is always even and the second always odd, the value of the both constituent functions in the residual mapping~(\ref{CommLett:Map_unique}) can be even or odd. However, in conjunction with $[\hat{x}]_{\rho/\tau}$, as explained below, it is possible to deduce from $M(\bar{\varepsilon}_{\rho/\tau})$ which of the constituent functions was used.
\ruimte

Consider the first constituent function of the residual mapping~(\ref{CommLett:Map_unique}).
\begin{eqnarray}
M(\bar{\varepsilon}_{\rho/\tau}) & = & \lfloor 2\bar{\varepsilon}_{\rho/\tau} \rfloor \nonumber \\
& = & \lfloor 2x - 2[\hat{x}]_{\rho/\tau} \rfloor \nonumber \\
& = & 2x - \left \lceil 2[\hat{x}]_{\rho/\tau} \right \rceil  \quad \because x \in \mathbb{Z}. \label{evenMeven} \end{eqnarray}
It follows from~(\ref{evenMeven}) that {\em the value of the first constituent function is even (odd) if $\lceil 2[\hat{x}]_{\rho/\tau} \rceil$ is even (odd)}. Now consider the second constituent function of~(\ref{CommLett:Map_unique})
\begin{eqnarray}
M(\bar{\varepsilon}_{\rho/\tau}) & = & - \lfloor 2\bar{\varepsilon}_{\rho/\tau} \rfloor -1 \nonumber \\
& = & -\lfloor 2x - 2[\hat{x}]_{\rho/\tau} \rfloor -1 \nonumber \\
& = & - (2x +1) + \left \lceil 2[\hat{x}]_{\rho/\tau} \right \rceil \quad \because x \in \mathbb{Z} \label{evenModd}
\end{eqnarray}
It follows from~(\ref{evenModd}) that {\em the value of the second constituent function is odd (even) if $\lceil 2[\hat{x}]_{\rho/\tau} \rceil$ is even (odd)}. These relationships between $\lceil 2[\hat{x}]_{\rho/\tau} \rceil$ and $M(\bar{\varepsilon}_{\rho/\tau})$ are summarized in Table~\ref{tab:relationship}.
\begin{table}[!bt]
\caption{The relationship between $\lceil 2[\hat{x}]_{\rho/\tau} \rceil$ and $M(\bar{\varepsilon}_{\rho/\tau})$}
\centering
\begin{tabular}{c | c | c c}
\hline \hline
& & \multicolumn{2}{c}{$M(\bar{\varepsilon}_{\rho/\tau})$} \\[0.5ex]
\hline
& & Even & Odd \\ [1ex]
\hline
& Even & $M(\bar{\varepsilon}_{\rho/\tau})= \lfloor 2\bar{\varepsilon}_{\rho/\tau}\rfloor$ & $M(\bar{\varepsilon}_{\rho/\tau})= - \lfloor 2\bar{\varepsilon}_{\rho/\tau} \rfloor -1$  \\[1ex]
\raisebox{3ex}{$\lceil 2[\hat{x}]_{\rho/\tau} \rceil$} & Odd & $M(\bar{\varepsilon}_{\rho/\tau})= - \lfloor 2\bar{\varepsilon}_{\rho/\tau} \rfloor -1$ & $M(\bar{\varepsilon}_{\rho/\tau})= \lfloor 2\bar{\varepsilon}_{\rho/\tau}\rfloor$ \\[1ex]
\hline
\end{tabular}
\label{tab:relationship}
\end{table}
These relationships now can be used to decide which of the constituent functions in the mapping~(\ref{CommLett:Map_unique}) was used. If both $\lceil 2[\hat{x}]_{\rho/\tau} \rceil$ and $M(\bar{\varepsilon}_{\rho/\tau})$ are even or both are odd then the first constituent function $M(\bar{\varepsilon}_{\rho/\tau})=\lfloor 2\bar{\varepsilon}_{\rho/\tau} \rfloor$ was used; otherwise the second constituent function $M(\bar{\varepsilon}_{\rho/\tau})=- \lfloor 2\bar{\varepsilon}_{\rho/\tau} \rfloor -1$ was used. Having decided on the constituent functions, the decoding now follows from~(\ref{evenMeven}) and~(\ref{evenModd}) as
\begin{equation}
x = \left \{ \begin{array}{ll} \frac{M(\bar{\varepsilon}_{\rho/\tau}) + \lceil 2[\hat{x}]_{\rho/\tau} \rceil}{2}, &  \textnormal{ if } M(\bar{\varepsilon}_{\rho/\tau}) + \lceil 2[\hat{x}]_{\rho/\tau} \rceil \textnormal{ is even}; \\ \frac{\lceil 2[\hat{x}]_{\rho/\tau} \rceil - M(\bar{\varepsilon}_{\rho/\tau}) - 1}{2}, & \textrm{otherwise.} \end{array} \right.
\label{eqn:DecisionTable}
\end{equation}
The pseudocode of the decoding algorithm is given in Algorithm~\ref{Algo-MRGDecode}.
\begin{algorithm}[!htb]
\caption{Decoding}
\label{Algo-MRGDecode}
\begin{algorithmic}
\STATE
\STATE {\bf Input}: $\hat{x} \in \mathbb{R}$, $\tau, \rho, m  \in \mathbb{Z}^+$, $j$, $k$
\STATE {\bf Output}: $x$
%\STATE
\STATE $M(\bar{\varepsilon}_{\rho/\tau}) := j m + k$;
\STATE $[\hat{x}]_{\rho/\tau} := \rho\lfloor \tau\hat{x}/\rho + 1/2 \rfloor /\tau$;\\
\IF{$M(\bar{\varepsilon}_{\rho/\tau})+ \lceil 2[\hat{x}]_{\rho/\tau} \rceil$ is even}
\STATE $x := \frac{\lceil 2[\hat{x}]_{\rho/\tau} \rceil  + M(\bar{\varepsilon}_{\rho/\tau})}{2}$;
\ELSE
\STATE $x := \frac{\lceil 2[\hat{x}]_{\rho/\tau} \rceil - M(\bar{\varepsilon}_{\rho/\tau}) - 1}{2}$;
\ENDIF
\end{algorithmic}
\end{algorithm}

\section{Analysis}
\label{MRGanalysis}
In Section~\ref{sec:asso-inte}, the association of non-negative integers with different residual intervals by the mapping $M(\bar{\varepsilon}_{\rho/\tau})$ is determined, which aids in computing the average code-length of the proposed coding scheme in Section~\ref{sec:modi-avg-len}.

\subsection{Association of non-negative integers to residual intervals}
\label{sec:asso-inte}
When operating at precision $\rho/\tau$, a real-valued prediction $\hat{x} \in [i + n \rho/\tau - \rho/2\tau,\, i + n \rho/\tau + \rho/2\tau)$, where $i, n \in \mathbb{Z}$ and $0 \leq n < \tau/\rho$, is rounded to
\begin{equation}
[\hat{x}]_{\rho/\tau} = i + n \rho/\tau.
\end{equation}
Then for a given $x$, the residual takes the form
\begin{eqnarray}
\bar{\varepsilon}_{\rho/\tau} & = & x - [\hat{x}]_{\rho/\tau} \nonumber \\
& = & (x-i) - n \rho/\tau.
\end{eqnarray}
Another relevant quantity is $\lceil 2[\hat{x}]_{\rho/\tau} \rceil = \lceil 2i +2n\rho/\tau \rceil$, which is even if either $n=0$ or $\tau/2\rho < n < \tau/\rho$ and odd if $0 < n \leq \tau/2\rho$. Now depending on $M(\bar{\varepsilon}_{\rho/\tau})$ and $\lceil 2[\hat{x}]_{\rho/\tau} \rceil$, there can be four cases.

\ruimte
{\em Case 1:} Both $\quad M(\bar{\varepsilon}_{\rho/\tau})$ and $\lceil 2[\hat{x}]_{\rho/\tau} \rceil$ are even

In this case, $M(\bar{\varepsilon}_{\rho/\tau}) = 2l$ for some $l \in \mathbb{Z}^+$ and either $n=0$ or $\tau/2\rho < n < \tau/\rho$. If $n=0$ then from Table~\ref{tab:relationship} it follows that
\begin{eqnarray}
& & \left \lfloor 2(x-i) - 2n \frac{\rho}{\tau} \right \rfloor  =  2l \nonumber \\
& \Rightarrow & 2(x - i) = 2l \quad \because n=0 \nonumber \\
& \Rightarrow & x = i + l.
\end{eqnarray}
Therefore, the non-negative integer associated with the residual $x - \hat{x} \in (l - \rho/2\tau, \, l + \rho/2\tau]$ is $2l$. For example, when $\rho=1$ and $\tau=4$, the non-negative integer associated with the interval $(-0.125, 0.125]$ is $2l=0$.

On the other hand, if $\tau/2\rho < n < \tau/\rho$ then from Table~\ref{tab:relationship} it follows that
\begin{eqnarray}
& & \left \lfloor 2(x-i) - 2n \frac{\rho}{\tau} \right \rfloor  =  2l \nonumber \\
& \Rightarrow & 2(x - i) -2 = 2l \quad \because \tau/2\rho < n < \tau/\rho \nonumber \\
& \Rightarrow & x = i + l+ 1.
\end{eqnarray}
Therefore, the non-negative integer associated with the residual $x - \hat{x} \in (l + 1 - n \rho/\tau - \rho/2\tau, \, l +1 - n \rho/\tau + \rho/2\tau]$ is $2l$ where $\tau/2\rho < n < \tau/\rho$. For example, when $\rho=1$ and $\tau=4$, we only have $n=3$ satisfying the condition $2 < n < 4$. Therefore, $2l=0$ must be associated with the interval $(0.125, 0.375]$.

\ruimte
{\em Case 2:} $\quad M(\bar{\varepsilon}_{\rho/\tau})$ is even and $\lceil 2[\hat{x}]_{\rho/\tau} \rceil$ is odd

In this case, $M(\bar{\varepsilon}_{\rho/\tau}) = 2l$ for some $l \in \mathbb{Z}^+$ and $0 < n \leq \tau/2\rho$. Now from Table~\ref{tab:relationship} it follows that
\begin{eqnarray}
& & - \left \lfloor 2(x-i) - 2n \frac{\rho}{\tau} \right \rfloor -1 =  2l \nonumber \\
& \Rightarrow & - 2(x - i) + 1 - 1 = 2l \quad \because 0 < n \leq \tau/2\rho \nonumber \\
& \Rightarrow & x = i - l.
\end{eqnarray}
Therefore, the non-negative integer associated with the residual $x - \hat{x} \in (-l - n \rho/\tau - \rho/2\tau, \, - l - n \rho/\tau + \rho/2\tau]$ is $2l$ where $0 < n \leq \tau/2\rho$. For example, when $\rho=1$ and $\tau=4$, we have $n=1$ and $n=2$ such that $0< n \leq 2$. Thus $2l=0$ must be associated with the interval $(-0.375, -0.125]$.

\ruimte
{\em Case 3:} $\quad M(\bar{\varepsilon}_{\rho/\tau})$ is odd and $\lceil 2[\hat{x}]_{\rho/\tau} \rceil$ is even

In this case, $M(\bar{\varepsilon}_{\rho/\tau}) = 2l+1$ for some $l \in \mathbb{Z}^+$ and either $n=0$ or $\tau/2\rho < n < \tau/\rho$. If $n=0$ then from Table~\ref{tab:relationship} it follows that
\begin{eqnarray}
& & - \left \lfloor 2(x-i) - 2n \frac{\rho}{\tau} \right \rfloor - 1  =  2l + 1 \nonumber \\
& \Rightarrow & - 2(x - i) - 1 = 2l + 1 \quad \because n=0 \nonumber \\
& \Rightarrow & x = i - l -1.
\end{eqnarray}
Therefore, the non-negative integer associated with the residual $x - \hat{x} \in (- l -1 - \rho/2\tau, \, - l -1 + \rho/2\tau]$ is $2l+1$. For example, when $\rho=1$ and $\tau=4$, the non-negative integer associated with the interval $(-1.125, -0.875]$ must be $2l+1=1$.

On the other hand, if $\tau/2\rho < n < \tau/\rho$ then from Table~\ref{tab:relationship} it follows that
\begin{eqnarray}
& & - \left \lfloor 2(x-i) - 2n \frac{\rho}{\tau} \right \rfloor - 1 =  2l+1 \nonumber \\
& \Rightarrow & - 2(x - i) + 2 -1 = 2l + 1\quad \because \tau/2\rho < n < \tau/\rho \nonumber \\
& \Rightarrow & x = i - l.
\end{eqnarray}
Therefore, the non-negative integer associated with the residual $x - \hat{x} \in (-l - n \rho/\tau - \rho/2\tau, \, - l - n \rho/\tau + \rho/2\tau]$ is $2l+1$ where $\tau/2\rho < n < \tau/\rho$. For example, when $\rho=1$ and $\tau=4$, we only have $n=0$ such that $2 < n=3 < 4$. Therefore, $2l+1=1$ must be associated with the interval $(-0.875, -0.625]$.

\ruimte
{\em Case 4:} Both $\quad M(\bar{\varepsilon}_{\rho/\tau})$ and $\lceil 2[\hat{x}]_{\rho/\tau} \rceil$ are odd

In this case, $M(\bar{\varepsilon}_{\rho/\tau}) = 2l+1$ for some $l \in \mathbb{Z}^+$ and $0 < n \leq \tau/2\rho$. Now from Table~\ref{tab:relationship} it follows that
\begin{eqnarray}
& & \left \lfloor 2(x-i) - 2n \frac{\rho}{\tau} \right \rfloor  =  2l+1 \nonumber \\
& \Rightarrow & 2(x - i) - 1 = 2l + 1 \quad \because 0 < n \leq \tau/2\rho \nonumber \\
& \Rightarrow & x = i + l + 1.
\end{eqnarray}
Therefore, the non-negative integer associated with the residual $x - \hat{x} \in (l+1 - n \rho/\tau - \rho/2\tau, \, l +1 - n \rho/\tau + \rho/2\tau]$ is $2l+1$ where $0 < n \leq \tau/2\rho$. For example, when $\rho=1$ and $\tau=4$, we have $n=1$ and $n=2$ satisfying $0< n \leq 2$, and thus $2l+1=1$ must be associated with the interval $(0.375, 0.875]$.

\begin{figure*}[!tb]
\centering
\includegraphics[width = 10.5cm]{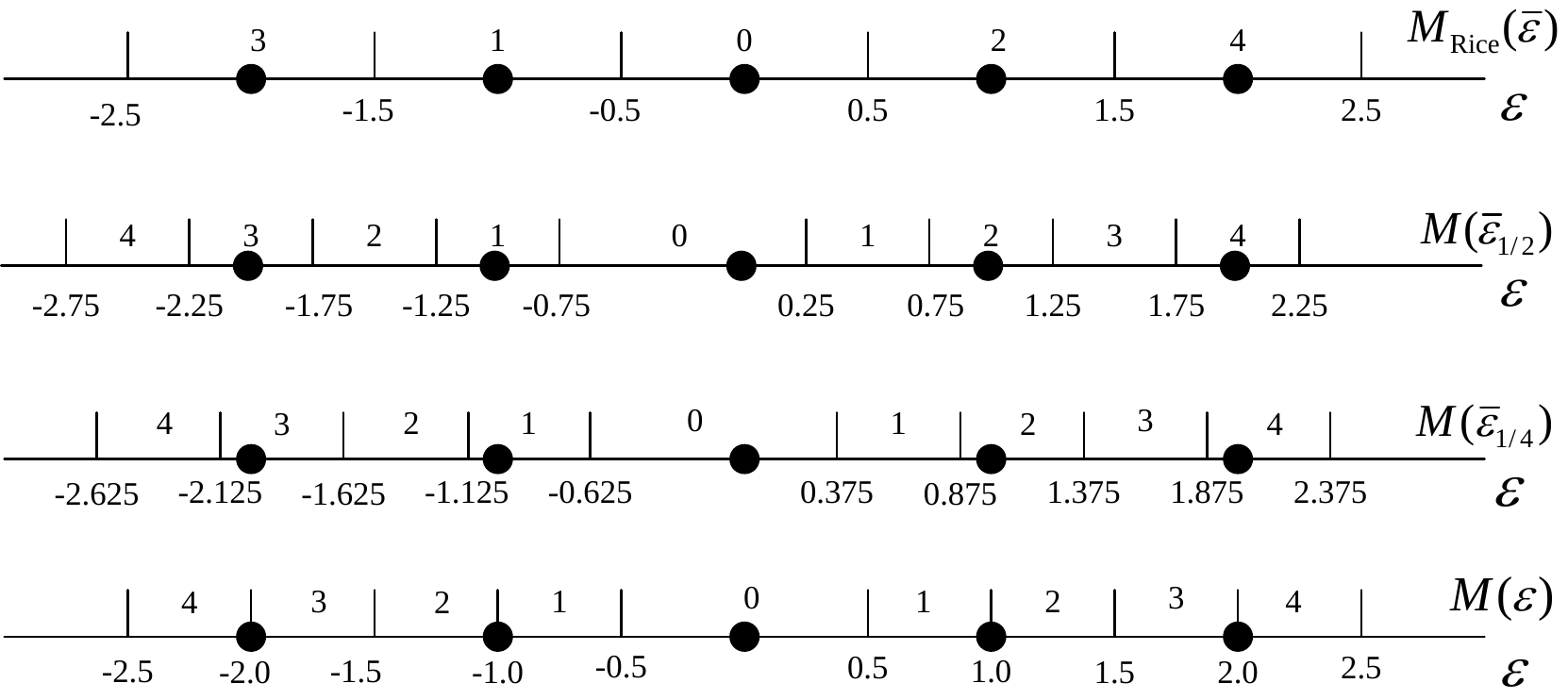}
\caption{ The assignment of non-negative integers to different residual intervals for different values of $\rho/\tau$. While at a coarse precision the assignment is asymmetric, at finer precision the assignment becomes symmetric.}
\label{fig:CommLetTau}
\end{figure*}

\ruimte
The assignment of non-negative integers by the mapping $M(\bar{\varepsilon}_{\rho/\tau})$ to different intervals of the residual $\varepsilon$ for different values of $\rho/\tau$ is depicted in Fig.~\ref{fig:CommLetTau}.

\subsection{Average code-length}
\label{sec:modi-avg-len}
It follows from Fig.~\ref{fig:CommLetTau} that when $\rho = \tau=1$, which corresponds to the Rice mapping~(\ref{eqn:riceMapping}), the assignment of non-negative integers to different residual intervals is asymmetric as equally probable, opposite-signed intervals are mapped to different integers. This asymmetry results in the assignments of codes of different lengths to equally probable residual intervals. However, this asymmetry reduces at finer precision and the assignment becomes symmetric in the asymptotic case of $\rho/\tau \to 0$.

\ruimte
The analysis of the modified Rice-Golomb code becomes simpler for the asymptotic case of $\rho/\tau \to 0$ due to the symmetric assignment of non-negative integers to the residual intervals. It can be observed from Fig.~\ref{fig:CommLetTau} that the assignment corresponding to $M(\bar{\varepsilon}_{\rho/\tau})$ is same as the assignment corresponding to $M(\varepsilon)$ but with a left shift of $\rho/2\tau$. When $M(\bar{\varepsilon}_{\rho/\tau})$ is encoded using a Golomb code with a parameter $m$, this left shift is also reflected in the association of code-lengths with different residual intervals. Let for  a given $m$, the length of the code associated with $\varepsilon$ in the modified Rice-Golomb code at precision $\rho/\tau$ be $\ell_m^{\rho/\tau}(\varepsilon)$. In the asymptotic case of $\rho/\tau \to 0$, the code-length $\ell_m^{\rho/\tau}(\varepsilon)$ will be denoted by $\ell_m(\varepsilon)$. Now let us consider two cases depending on whether $m$ is a power of $2$ or not.

\subsubsection{The case $m = 2^\beta$}
\begin{figure*}[!tb]
\centering
\includegraphics[width=14cm]{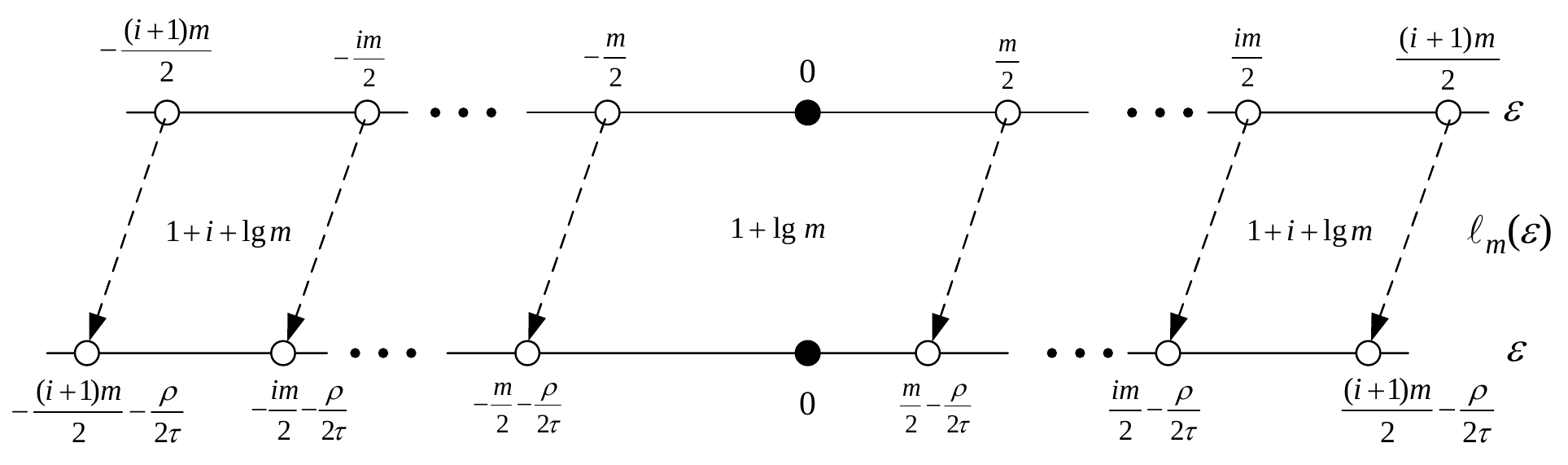}
\caption{Association of code-lengths to different residual intervals when $m = 2^\beta, \beta \in \mathbb{N}$. The code-length assignment shifts to the left by $\rho/2\tau$ as compared to the symmetric assignment achieved at precision $\rho/\tau \rightarrow 0$.}
\label{fig:Ch5CaseILeftShift}
\end{figure*}

When $m = 2^\beta$, the minimal binary representation of $k$ always takes $\lg m$ bits. In the asymptotic case of $\rho/\tau \rightarrow 0$, for any residual $\varepsilon$ such that $im/2 \leq |\varepsilon|< (i+1)m/2$, we get $j = i$. As the unary representation of $i$ requires $i+1$ bits, the length of the code associated with $\varepsilon$ is $\ell_{m}(\varepsilon) = 1 + i + \lg m$ . Now it follows from Fig.~\ref{fig:Ch5CaseILeftShift} that the code-length associated with the left-shifted intervals $(im/2-\rho/2\tau, (i+1)m/2-\rho/2\tau)$ and $(-(i+1)m/2-\rho/2\tau, -im/2-\rho/2\tau)$ is $\ell_{m}(\varepsilon) = 1 + i + \lg m$.
This observation can be used to determine the average code-length of the modified Rice-Golomb code. Given $m$, let the average code-length achieved with the modified Rice-Golomb code, in encoding residuals that are Laplace distributed with parameter $\theta$, be $L_m^{\rho/\tau}(\theta)$. Then we have the following theorem.
\begin{thm}
If $m = 2^\beta, \beta \in \mathbb{N}$, then $L_m^{\rho/\tau}(\theta) = 1 + \lg m + \frac{1}{2}\frac{\theta^{m/2}}{1 - \theta^{m/2}}(\theta^{\frac{\rho}{2\tau}} + \theta^{-\frac{\rho}{2\tau}}).$
\label{theo-AvgLenPO2MRG}
\end{thm}
\pf
%See appendix~\ref{app:avglen1}
We have
\begin{eqnarray}
L_m^{\rho/\tau}(\theta) & = & \int_{-\infty}^{+\infty} \ell_m^{\rho/\tau}(\varepsilon) f_\theta(\varepsilon) d\varepsilon \nonumber \\
& = & \int_{-\infty}^{0 } \ell_m^{\rho/\tau}(\varepsilon) f_\theta(\varepsilon) d\varepsilon + \int_{0}^{+\infty} \ell_m^{\rho/\tau}(\varepsilon) f_\theta(\varepsilon) d\varepsilon \nonumber \\
& = & L_m^{\rho/\tau-}(\theta) + L_m^{\rho/\tau+}(\theta) \label{eqn:SumOfTwo}
\end{eqnarray}
where $L_m^{\rho/\tau-}(\theta) = \int_{-\infty}^{0 } \ell_m^{\rho/\tau}(\varepsilon) f_\theta(\varepsilon) d\varepsilon$ and $ L_m^{\rho/\tau+}(\theta) = \int_{0}^{+\infty} \ell_m^{\rho/\tau}(\varepsilon) f_\theta(\varepsilon) d\varepsilon$. Now it follows from Fig.~\ref{fig:Ch5CaseILeftShift} that
\begin{equation}
L_m^{\rho/\tau-}(\theta) = -\frac{\ln \theta}{2} \int_{-\frac{m}{2}- \frac{\rho}{2\tau}}^{0}(1 + \lg m)\theta^{-\varepsilon} d\varepsilon -\frac{\ln \theta}{2} \sum_{i=1}^{\infty}\int_{-\frac{(i+1)m}{2} - \frac{\rho}{2\tau}}^{-\frac{im}{2} - \frac{\rho}{2\tau}}(1 + i + \lg m)\theta^{-\varepsilon} d\varepsilon.
\end{equation}
Substituting $\varepsilon = -\varepsilon$,
\begin{equation}
L_m^{\rho/\tau-}(\theta)  = -\frac{\ln \theta}{2} \int_{0}^{\frac{m}{2}+ \frac{\rho}{2\tau}} (1 + \lg m)\theta^{\varepsilon} d\varepsilon -\frac{\ln \theta}{2} \sum_{i=1}^{\infty}\int_{\frac{im}{2} + \frac{\rho}{2\tau}}^{\frac{(i+1)m}{2} + \frac{\rho}{2\tau}} (1 + i + \lg m) \theta^{\varepsilon} d\varepsilon.
\end{equation}
After algebraic simplification,
\begin{equation}
L_m^{\rho/\tau-}(\theta)  = \frac{1}{2}(1 + \lg m) + \frac{1}{2}\frac{\theta^{m/2}}{1-\theta^{m/2}}\theta^{\rho/2\tau}.
\label{eqn:NegOfTwo}
\end{equation}
From Fig.~\ref{fig:Ch5CaseILeftShift} it also follows that,
\begin{equation}
L_m^{\rho/\tau+}(\theta) = -\frac{\ln \theta}{2} \int_{0}^{\frac{m}{2}- \frac{\rho}{2\tau}}(1 + \lg m)\theta^{\varepsilon} d\varepsilon -\frac{\ln \theta}{2} \sum_{i=1}^{\infty}\int_{\frac{im}{2} - \frac{\rho}{2\tau}}^{\frac{(i+1)m}{2} - \frac{\rho}{2\tau}}(1 + i + \lg m)\theta^{\varepsilon} d\varepsilon.
\end{equation}
After algebraic simplification,
\begin{equation}
L_m^{\rho/\tau+}(\theta)  = \frac{1}{2}(1 + \lg m) + \frac{1}{2}\frac{\theta^{m/2}}{1-\theta^{m/2}}\theta^{-\rho/2\tau}.
\label{eqn:PosOfTwo}
\end{equation}
Now from (\ref{eqn:SumOfTwo}), (\ref{eqn:NegOfTwo}), and (\ref{eqn:PosOfTwo}) we get,
\begin{equation*}
L_m^{\rho/\tau}(\theta) = 1 + \lg m + \frac{1}{2}\frac{\theta^{m/2}}{1-\theta^{m/2}}(\theta^{\rho/2\tau} + \theta^{-\rho/2\tau}).
\end{equation*}
\eind

\subsubsection{The case $m \neq 2^\beta$}
First we determine the code-lengths associated with different intervals for the asymptotic case of $\rho/\tau \to 0$. Let $T = 2^{\lceil \lg m \rceil} - m - 1$. When $m \neq 2^\beta$, the minimal binary representation of $k$ takes $\lfloor \lg m \rfloor$ bits if $0 \leq k \leq T$ or $\lceil \lg m \rceil$ bits if $T+1 \leq k < m$. Thus the minimum possible code-length in this scheme is $1 + \lfloor \lg m \rfloor$ which results when $j=0$ and $0 \leq k \leq T$. Since $T+1<m$, this minimum code-length is associated with the interval $(-(T+1)/2, (T+1)/2)$. Beyond the point $|\varepsilon| = (T+1)/2$, we have $k>T$ and the minimal binary representation of $k$ requires $\lceil \lg m \rceil$ bits. However, the value of $j$ remains $0$ for $|\varepsilon|< m/2$. Therefore, $\ell_{m}(\varepsilon) = 1 + \lceil \lg m \rceil$ for $(T+1)/2 \leq |\varepsilon| < m/2$. Now, when $|\varepsilon| \geq m/2$ although $j$ becomes $1$ whose unary representation requires two bits, the value of $k$ becomes~$\leq T$ for $m/2 \leq |\varepsilon| < m/2 + (T+1)/2$. Hence, $\ell_{m}(\varepsilon) = 2 + \lfloor \lg m \rfloor$ for $m/2 \leq |\varepsilon| < m/2 + (T+1)/2$. Since, when $m \neq 2^\beta$, we have $\lceil \lg m \rceil = 1+\lfloor \lg m \rfloor$, for $(T+1)/2 \leq |\varepsilon| < m/2 + (T+1)/2$, we get $\ell_{m}(\varepsilon) = 2 + \lfloor \lg m \rfloor$. This calculation is generalized in the following lemma.
\begin{lem}
Given $m \neq 2^\beta$ for any integer $\beta \geq 0$, if $\frac{T+1}{2} + \frac{im}{2} \leq \varepsilon < \frac{T+1}{2} + \frac{(i+1)m}{2}$ then the code-length associated with $\varepsilon$ is $\ell_{m}(\varepsilon) = 2 + i + \lfloor \lg m \rfloor$.
\label{lem:interval}
\end{lem}
\pf
See Appendix~\ref{app:interval1}.
\eind
\pfend
\begin{figure*}[!tb]
\centering
\includegraphics[width=15cm]{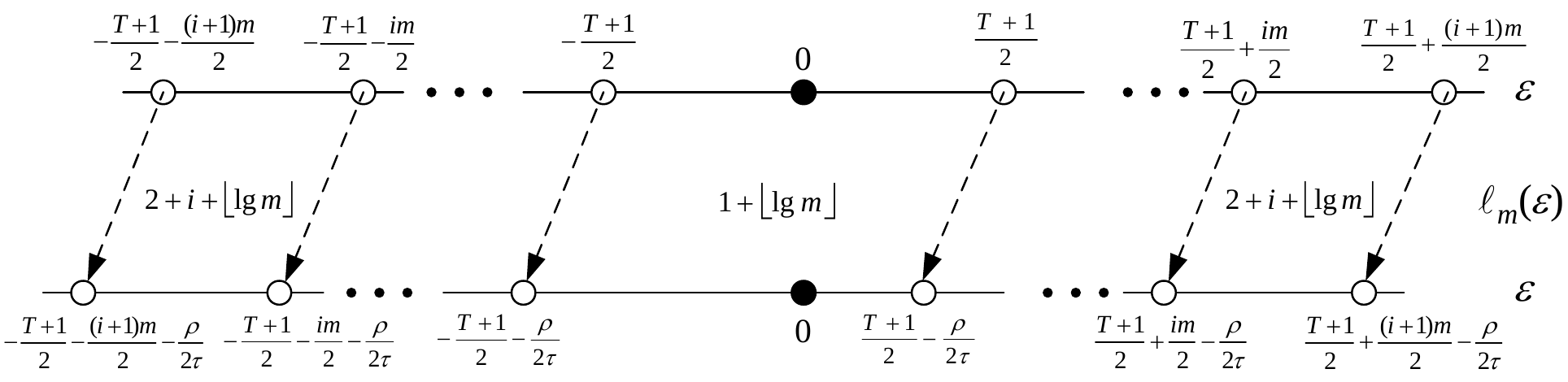}
\caption{Association of code-lengths to different residual intervals when $m \neq 2^\beta$ for any integer $\beta \geq 0$. The code-length assignment shifts to the left by $\rho/2\tau$ as compared to the symmetric assignment achieved at precision $\rho/\tau \rightarrow 0$.}
\label{fig:Ch5CaseIILeftShift}
\end{figure*}
Now for an arbitrary $\rho/\tau$, the minimum code-length $1 + \lfloor \lg m \rfloor$ is associated with the left-shifted interval $(-(T+1)/2-\rho/2\tau, (T+1)/2 - \rho/2\tau)$ and the code-length $2 + i +  \lfloor \lg m \rfloor$ is associated with the left-shifted intervals $((T+1)/2 + im/2-\rho/2\tau, (T+1)/2(i+1)m/2 - \rho/2\tau)$ and $(-(T+1)/2-(i+1)m/2 - \rho/2\tau, -(T+1)/2-im/2 - \rho/2\tau)$. Using these association of code-lengths to different intervals as depicted in Fig.~\ref{fig:Ch5CaseIILeftShift}, we can determine the average code-length $L_m^{\rho/\tau}(\theta)$.
\begin{thm}
If $m \neq 2^\beta$ for any integer $\beta \geq 0$ then $L_m^{\rho/\tau}(\theta) = 1 + \lfloor \lg m \rfloor + \frac{1}{2}\frac{\theta^{(T+1)/2}}{1 - \theta^{m/2}}(\theta^{\frac{\rho}{2\tau}} + \theta^{-\frac{\rho}{2\tau}}).$
\label{theo-AvgLenNPO2MRG}
\end{thm}
\pf
We have
\begin{eqnarray}
L_m^{\rho/\tau}(\theta) & = & \int_{-\infty}^{+\infty} \ell_m^{\rho/\tau}(\varepsilon) f_\theta(\varepsilon) d\varepsilon \nonumber \\
& = & \int_{-\infty}^{0 } \ell_m^{\rho/\tau}(\varepsilon) f_\theta(\varepsilon) d\varepsilon + \int_{0}^{+\infty} \ell_m^{\rho/\tau}(\varepsilon) f_\theta(\varepsilon) d\varepsilon \nonumber \\
& = & L_m^{\rho/\tau-}(\theta) + L_m^{\rho/\tau+}(\theta) \label{eqn:NOPSumOfTwo}
\end{eqnarray}
where $L_m^{\rho/\tau-}(\theta)  = \int_{-\infty}^{0 } \ell_m^{\rho/\tau}(\varepsilon) f_\theta(\varepsilon) d\varepsilon$ and $L_m^{\rho/\tau+}(\theta) = \int_{0}^{+\infty} \ell_m^{\rho/\tau}(\varepsilon) f_\theta(\varepsilon) d\varepsilon$. Now it follows from Fig.~\ref{fig:Ch5CaseIILeftShift} that
\begin{equation}
L_m^{\rho/\tau-}(\theta) = -\frac{\ln \theta}{2} \int_{-\frac{(T+1)}{2}- \frac{\rho}{2\tau}}^{0}(1 + \lfloor \lg m \rfloor )\theta^{-\varepsilon} d\varepsilon -\frac{\ln \theta}{2} \sum_{i=0}^{\infty}\int_{-\frac{(T+1)}{2} -\frac{(i+1)m}{2} - \frac{\rho}{2\tau}}^{-\frac{(T+1)}{2} -\frac{im}{2} - \frac{\rho}{2\tau}}(2 + i + \lfloor \lg m \rfloor)\theta^{-\varepsilon} d\varepsilon.
\end{equation}
Substituting $\varepsilon = -\varepsilon$,
\begin{equation}
L_m^{\rho/\tau-}(\theta) = -\frac{\ln \theta}{2} \int_{0}^{\frac{(T+1)}{2}+ \frac{\rho}{2\tau}} (1 + \lfloor \lg m \rfloor )\theta^{\varepsilon} d\varepsilon -\frac{\ln \theta}{2} \sum_{i=0}^{\infty}\int_{\frac{(T+1)}{2} +\frac{im}{2} + \frac{\rho}{2\tau}}^{\frac{(T+1)}{2} +\frac{(i+1)m}{2} + \frac{\rho}{2\tau}}(2 + i + \lfloor \lg m \rfloor)\theta^{\varepsilon} d\varepsilon.
\end{equation}
After algebraic simplification,
\begin{equation}
L_m^{\rho/\tau-}(\theta) = \frac{1}{2}(1 + \lfloor \lg m \rfloor) + \frac{1}{2}\frac{\theta^{(T+1)/2}}{1-\theta^{m/2}}\theta^{\rho/2\tau}.
\label{eqn:NOPNegOfTwo}
\end{equation}
From Fig.~\ref{fig:Ch5CaseIILeftShift} it also follows that,
\begin{equation}
L_m^{\rho/\tau+}(\theta) = -\frac{\ln \theta}{2} \int_{0}^{\frac{(T+1)}{2}- \frac{\rho}{2\tau}} (1 + \lfloor \lg m \rfloor )\theta^{\varepsilon} d\varepsilon -\frac{\ln \theta}{2} \sum_{i=0}^{\infty}\int_{\frac{(T+1)}{2} +\frac{im}{2} - \frac{\rho}{2\tau}}^{\frac{(T+1)}{2} +\frac{(i+1)m}{2} - \frac{\rho}{2\tau}}(2 + i + \lfloor \lg m \rfloor)\theta^{\varepsilon} d\varepsilon.
\end{equation}
After algebraic simplification,
\begin{equation}
L_m^{\rho/\tau+}(\theta) = \frac{1}{2}(1 + \lfloor \lg m \rfloor) + \frac{1}{2}\frac{\theta^{(T+1)/2}}{1-\theta^{m/2}}\theta^{-\rho/2\tau}.
\label{eqn:NOPPosOfTwo}
\end{equation}
Now from (\ref{eqn:NOPSumOfTwo}), (\ref{eqn:NOPNegOfTwo}), and (\ref{eqn:NOPPosOfTwo}) we get,
\begin{equation*}
L_m^{\rho/\tau}(\theta) = 1 + \lfloor \lg m \rfloor + \frac{1}{2}\frac{\theta^{(T+1)/2}}{1-\theta^{m/2}}(\theta^{\rho/2\tau} + \theta^{-\rho/2\tau})
\end{equation*}
\eind
\pfend
From Theorem~\ref{theo-AvgLenPO2MRG} and Theorem~\ref{theo-AvgLenNPO2MRG}, the average code-length of the modified Rice-Golomb code can be summarized~as,
\begin{equation}
L_m^{\rho/\tau}(\theta) = \left \{ \begin{array}{ll} 1 + \lg m + \frac{1}{2} \frac {\theta^{\frac {m}{2}}}{1 - \theta^{\frac {m}{2}}}(\theta^{\rho/2\tau} + \theta^{-\rho/2\tau}), \textnormal{ if } m = 2^{\beta}, \beta \in \mathbb{N}; \\ 1 + \lfloor \lg m \rfloor + \frac{1}{2}\frac {\theta^{\frac{2^{\lceil \lg m \rceil}-m }{2}}}{1 - \theta^{\frac{m}{2}}}(\theta^{\rho/2\tau} + \theta^{-\rho/2\tau}), \quad \textnormal{otherwise.} \end{array} \right.
\label{eqn:sumMRG}
\end{equation}
Let denote the asymptotic average code-length with $L_m(\theta)$. Then it is immediate from~(\ref{eqn:sumMRG}) that
\begin{equation}
L_m(\theta) = \left \{ \begin{array}{ll} 1 + \lg m +  \frac {\theta^{\frac {m}{2}}}{1 - \theta^{\frac {m}{2}}}, \quad \textnormal{ if }m = 2^{\beta}, \beta \in \mathbb{N}; \\
1 + \lfloor \lg m \rfloor + \frac {\theta^{\frac{2^{\lceil \lg m \rceil}-m }{2}}}{1 - \theta^{\frac{m}{2}}}, \quad \textnormal{otherwise.} \end{array} \right.
\label{eqn:asySumMRG}
\end{equation}

\subsection{Optimal value of $m$}
We first determine the association between $\theta$ and the optimal value of $m$ for the asymptotic case of $\rho/\tau \to 0$. The relation between the optimal value of $m$ and $\theta$ is summarized in the following theorem.
\begin{thm}
\label{thm:optm}
When $\rho/\tau \to 0$, the value of $m$ is optimal for $\theta \in [\phi^2_{m-1}, \, \phi^2_{m}]$ where $\phi_m$ is the unique root of $f_m(\phi) = \phi^{m+1} + \phi^m - 1$ in the interval $(0,1)$.
\end{thm}
Before proving the theorem, we present few lemmas that will aid in the proof of the theorem.
%%%%%%%%%%%%%%%%%%%%%%%%%%%%%%%%%%%%%%%%%%%%%%%%%%%%%%%%%%%%%%%%%%
\begin{lem}
\label{lem:unique}
$f_n(\phi) = \phi^{n+1} + \phi^n-1 = 0$, $n>0$, has a unique root in $(0, 1)$.
\end{lem}
\pf
See Appendix~\ref{proof:lem:unique}.
\pfend
%%%%%%%%%%%%%%%%%%%%%%%%%%%%%%%%%%%%%%%%%%%%%%%%%%%%%%%%%%%%%%%%%%
\begin{lem}
$\phi^2_{m}$ is a unique root of $L_m(\theta) = L_{m+1}(\theta)$ in $(0, 1)$, where $\phi_{m}$ is the unique root of $\phi^{m+1} + \phi^{m} - 1=0$ in $(0, 1)$.
\label{lem:intersect}
\end{lem}
\pf
See Appendix~\ref{proof:lem:intersect}.
\pfend
%%%%%%%%%%%%%%%%%%%%%%%%%%%%%%%%%%%%%%%%%%%%%%%%%%%%%%%%%%%%%%%%%%
\begin{lem}
$\phi_{m+1} > \phi_{m}$
\label{lem:mc1GEQmc}
\end{lem}
\pf
See Appendix~\ref{proof:lem:mc1GEQmc}.
\pfend
%%%%%%%%%%%%%%%%%%%%%%%%%%%%%%%%%%%%%%%%%%%%%%%%%%%%%%%%%%%%%%%%%%
\begin{lem}
$\lim_{\theta \to 0} L_{m+1}(\theta) \geq \lim_{\theta \to 0} L_m(\theta).$
\label{lem:fromabove}
\end{lem}
\pf
See Appendix~\ref{proof:lem:fromabove}.
\pfend
%%%%%%%%%%%%%%%%%%%%%%%%%%%%%%%%%%%%%%%%%%%%%%%%%%%%%%%%%%%%%%%%%%
\begin{figure*}[!tb]
\centering
\includegraphics[width=7.5cm]{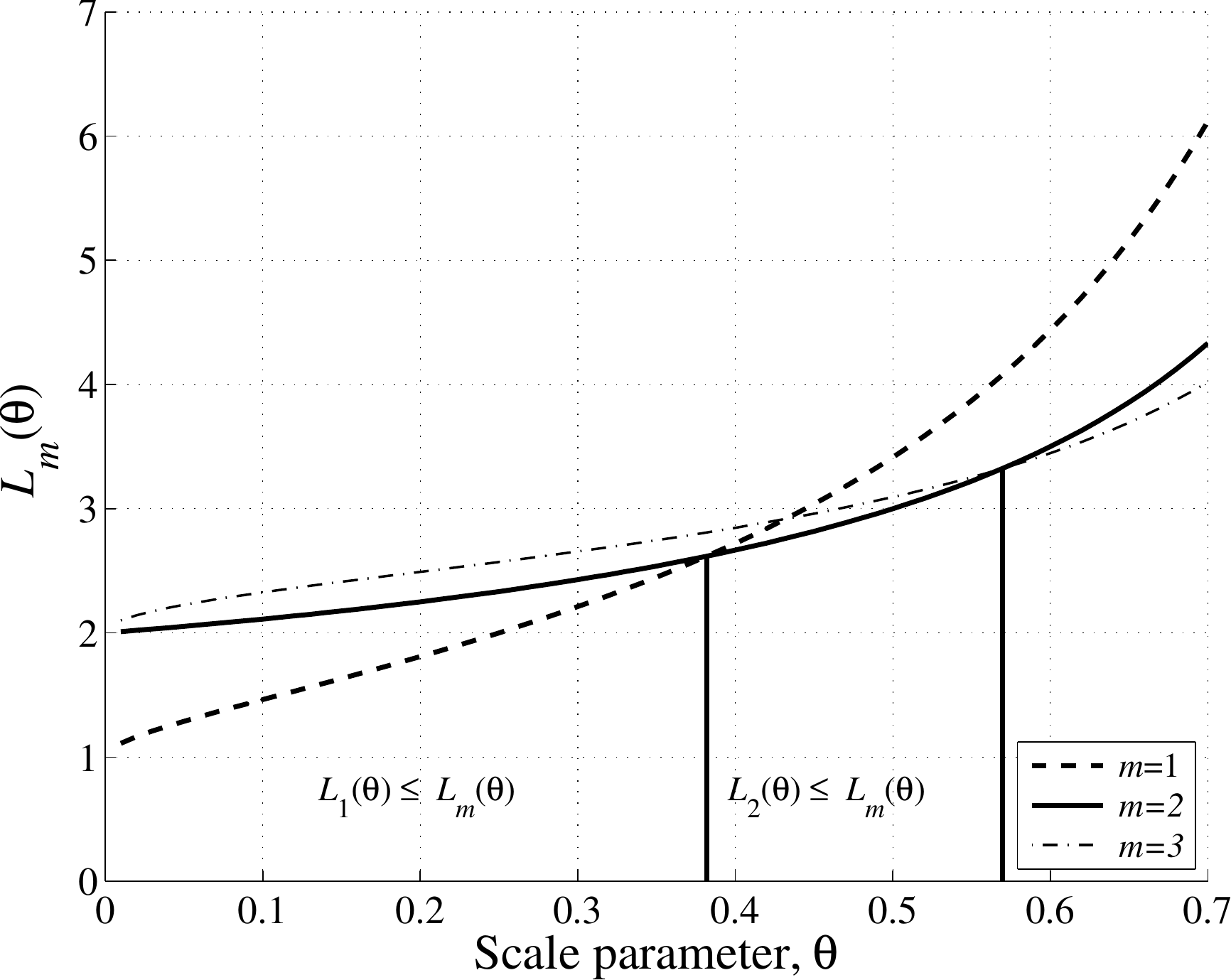}
\caption{$L_1(\theta)$ intersects $L_2(\theta)$ at point $\phi^2_1 = 0.3820$ from above and $L_2(\theta)$ intersects $L_3(\theta)$ at point $\phi^2_2 = 0.5698$ from above. Therefore, $L_1(\theta) \leq L_m(\theta)$ for all $m \neq 1$ and $m=1$ is optimal for $\theta \in (0, 0.3820]$. Similarly, $m=2$ is optimal for $\theta \in [0.3820, 0.5698]$.}
\label{fig:OptimaMc}
\end{figure*}
Now we prove Theorem~\ref{thm:optm}.
\pf
By Lemma~\ref{lem:intersect}, $L_{m-1}(\theta)$ and $L_m(\theta)$ intersect only at one point $\phi_{m-1}^2$ in $(0, 1)$. According to Lemma~\ref{lem:fromabove},  $L_m(\theta)$ intersects $L_{m-1}(\theta)$ from above at point $\phi_{m-1}^2$ (see Fig.~\ref{fig:OptimaMc}). Therefore,
\begin{equation}
L_m(\theta) < L_{m-1}(\theta), \textnormal{ for } \theta > \phi_{m-1}^2.
\end{equation}
Similarly, $L_{m+1}(\theta)$ intersects $L_m(\theta)$ from above at point $\phi_{m}^2$ (see Fig.~\ref{fig:OptimaMc}) and
\begin{equation}
L_{m+1}(\theta) < L_m(\theta), \textnormal{ for } \theta > \phi_{m}^2.
\end{equation}
Now $\phi_{m} > \phi_{m-1}$ according to Lemma~\ref{lem:mc1GEQmc} and thus $L_m(\theta) < L_{m^\prime}(\theta)$ in the interval $\phi^2_{m-1} \leq \theta \leq \phi^2_{m}$   for all  $m^\prime \neq m$.
\eind
\pfend

Although the above theorem states the relationship between the optimal value of $m$ and the scale parameter $\theta$, the determination of the optimal value of $m$, for a given value of $\theta$, is not straightforward. Therefore, we propose the following indirect method. Given $m$, the interval of $\theta$ in which $m$ is optimal is determined and stored in a table. Then for a given value of $\theta$, finding the optimal value of $m$ will involve a table look up. Since optimal value of $m$ increases with $\theta$, the logarithmic time binary search algorithm~\cite{ArtMathKnuth} can be used for searching the table. For example, if $32$ intervals corresponding to $32$ values of $m$ are stored in the table, only five comparisons are required to determine the optimal value of $m$ for a given $\theta$. In practice, however, only a few intervals need to be stored. For instance, $m=32$ is optimal for $\theta \in (0.9782, 0.9789)$, which corresponds to an almost uniform distribution. However, if the predictions are good enough, the distribution will be highly peaked at zero and the value of $\theta$ will be small. Therefore, if only the first few intervals are stored in the table and the maximum value of $m$ in the table is used when $\theta$ falls outside the range of the table, there will be negligible loss of compression efficiency.

\ruimte
Although a relationship between $\theta$ and the optimal value of $m$ exists for the asymptotic case of $\rho/\tau \rightarrow 0$~(Theorem~\ref{thm:optm}), no such closed form expression is readily available for an arbitrary value of $\rho/\tau$. For a given $\theta$, let $m_\theta$ be the optimal value of $m$ at the asymptotic precision of $\rho/\tau \rightarrow 0$. Now we can demonstrate that, if the coder operates at precision $\rho/\tau$ and uses $m_\theta$ as the coding parameter, then it incurs a negligible redundancy when $\rho/\tau$ is sufficiently small. When operating at precision $\rho/\tau$, let the sub-optimal use of $m_\theta$ results in a redundancy of $\Delta^{\rho/\tau}(\theta)$ bits as compared to $L_{m_\theta}(\theta)$. Then it follows from~(\ref{eqn:sumMRG}) and~(\ref{eqn:asySumMRG}) that
\begin{equation}
\Delta^{\rho/\tau}(\theta)= \left \{ \begin{array}{ll} \frac {\theta^{\frac {m_\theta}{2}}}{1 - \theta^{\frac {m_\theta}{2}}}(\theta^{\rho/2\tau} + \theta^{-\rho/2\tau}-2), \textnormal{ if } m_\theta = 2^{\beta}, \beta \in \mathbb{N}; \\ \frac {\theta^{\frac{2^{\lceil \lg m_\theta \rceil}-m_\theta }{2}}}{1 - \theta^{\frac{m_\theta}{2}}}(\theta^{\rho/2\tau} + \theta^{-\rho/2\tau}-2), \quad \textnormal{otherwise.} \end{array} \right.
\label{eqn:redunSub}
\end{equation}

\begin{table}
	\centering
	\caption{The maximum redundancy incurred by the sub-optimal use of $m_\theta$ for $\theta = 0.01,0.02, \ldots,  0.97$ at different precisions.}
		\begin{tabular}{c|cccccc} \hline
			& \multicolumn{6}{c}{Precision, $\rho/\tau$} \\
			& 4/5 & 1/2 & 1/4 & 1/8 & 1/16 & 1/32 \\ \hline
			Maximum redundancy(\%)	& 22.34 &	7.39 &	1.72 &	0.42 &	0.11 & 0.03 \\ \hline
		\end{tabular}
	\label{table:redunSub}
\end{table}
%$\Max \{ \Delta^{\rho/\tau}(\theta), {\theta = 0.01, \ldots, 0.97} \}$
The maximum redundancy incurred by the sub-optimal use of $m_\theta$ as compared to $L_{m_\theta}(\theta)$ at different precisions, is shown in Table~\ref{table:redunSub}. It follows from the table that, although at coarse precision this sub-optimal strategy results in significant redundancy, at finer precision the maximum redundancy becomes negligible. For example, at precision $1/16$, the maximum redundancy is bounded below $0.11\%$.

\subsection{Empirical estimation of $\theta$ and optimal value of $m$}
Besides having superior performance, the proposed scheme has several advantages over the Rice-Golomb code when simplicity of implementation is of concern. The Rice-Golomb code being asymmetric is difficult to analyze. There are no known method for the determination of the optimal choice of $m_g$, for given a $\theta$. Even the evaluation of the average code-length is difficult for the Rice-Golomb code. The Rice-Golomb code, first proposed in~\cite{RiceMapping}, divides the data sequence into blocks of length $J$ and computes the code-lengths using a set of values for the parameter $m_g$. The value of $m_g$ that yields the shortest code-length is then selected. This approach involves a delay at the beginning of the transmission of each block and thus is not suitable for on line symbol-by-symbol encoding. Moreover, Rice-Golomb code also requires the explicit transmission of the chosen parameter value $m_g$. In contrast, the $\theta$ of a Laplace distribution can be estimated on line from previously encoded sequences $x_0, x_1, \ldots, x_{t-1}$ and the corresponding predictions $\hat{x}_0, \hat{x}_1, \ldots, \hat{x}_{t-1}$, as $\hat{\theta} = e^{-t/S_t}$ where $S_t = \sum_0^{t-1}|x_i - \hat{x}_i| = \sum_0^{t-1}|\varepsilon_i|$ (see~\cite{Krishna05}). From the estimated $\hat{\theta}$, the optimal value of $m$ can be determined using the strategy described in the previous sub-section. Since, the decoder also has the same previously encoded data and the corresponding predictions, it can also estimate the same $\hat{\theta}$ and determine the same value of $m$.

\section{Comparative Analysis and Experimental Results}
\label{sec:exp}
In this section we first demonstrate that the proposed coding scheme at precision $\rho/\tau \to 0$ always outperforms the Rice-Golomb code. Then we presents simulation results to analyze the performance of the proposed scheme at different values of the scale parameter $\theta$ and precision $\rho/\tau$.
%Finally, we empirically show that given $\theta$ and $\rho/\tau$, instead of the using the optimal parameter $m$, use of $m^*$, which is optimal for $\rho/\tau \to 0$,  results in negligible performance loss.
\ruimte
\begin{figure}%
\centering
\subfloat[][]{
\includegraphics[width=7.6cm]{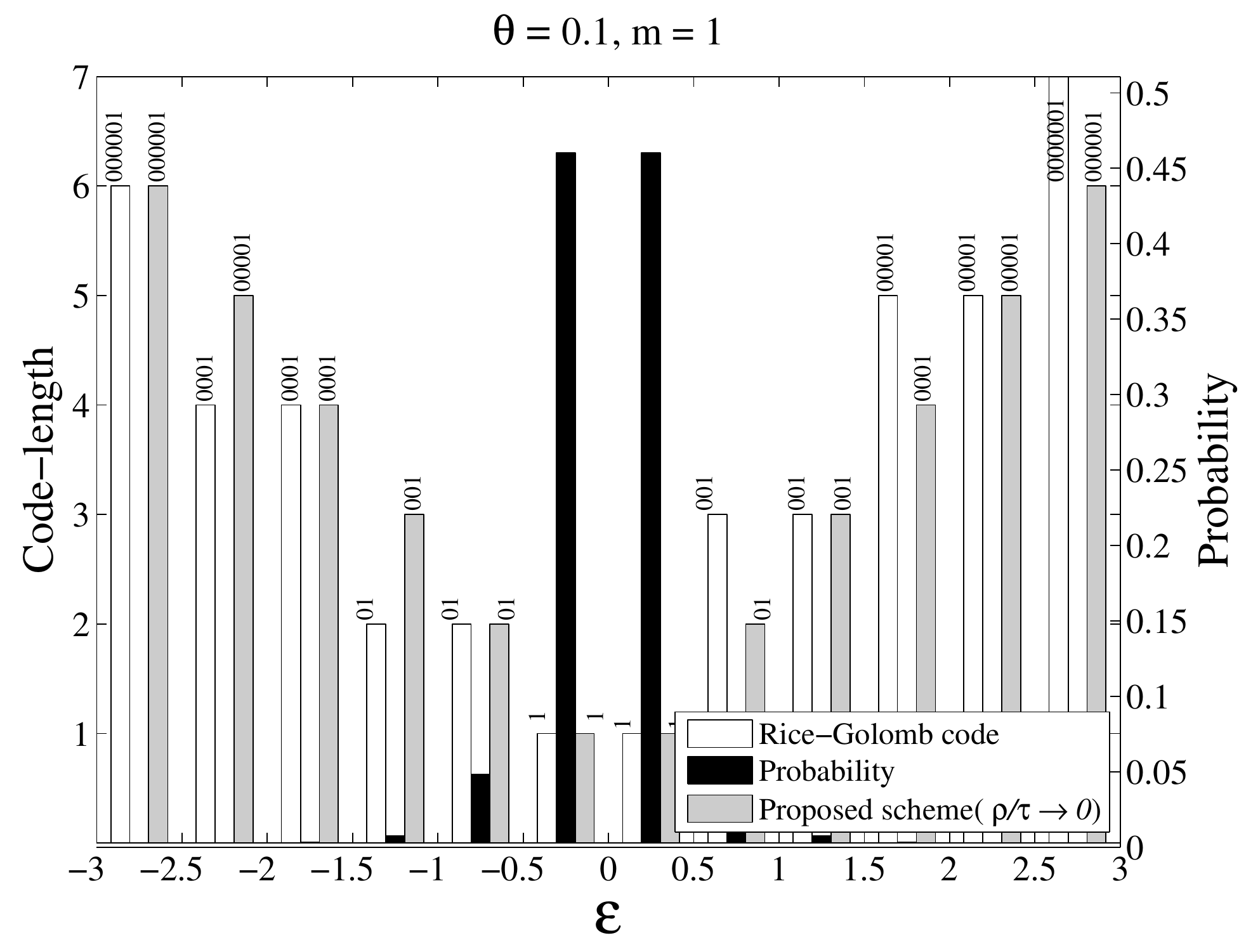}
} \\
\subfloat[][]{
\includegraphics[width=7.6cm]{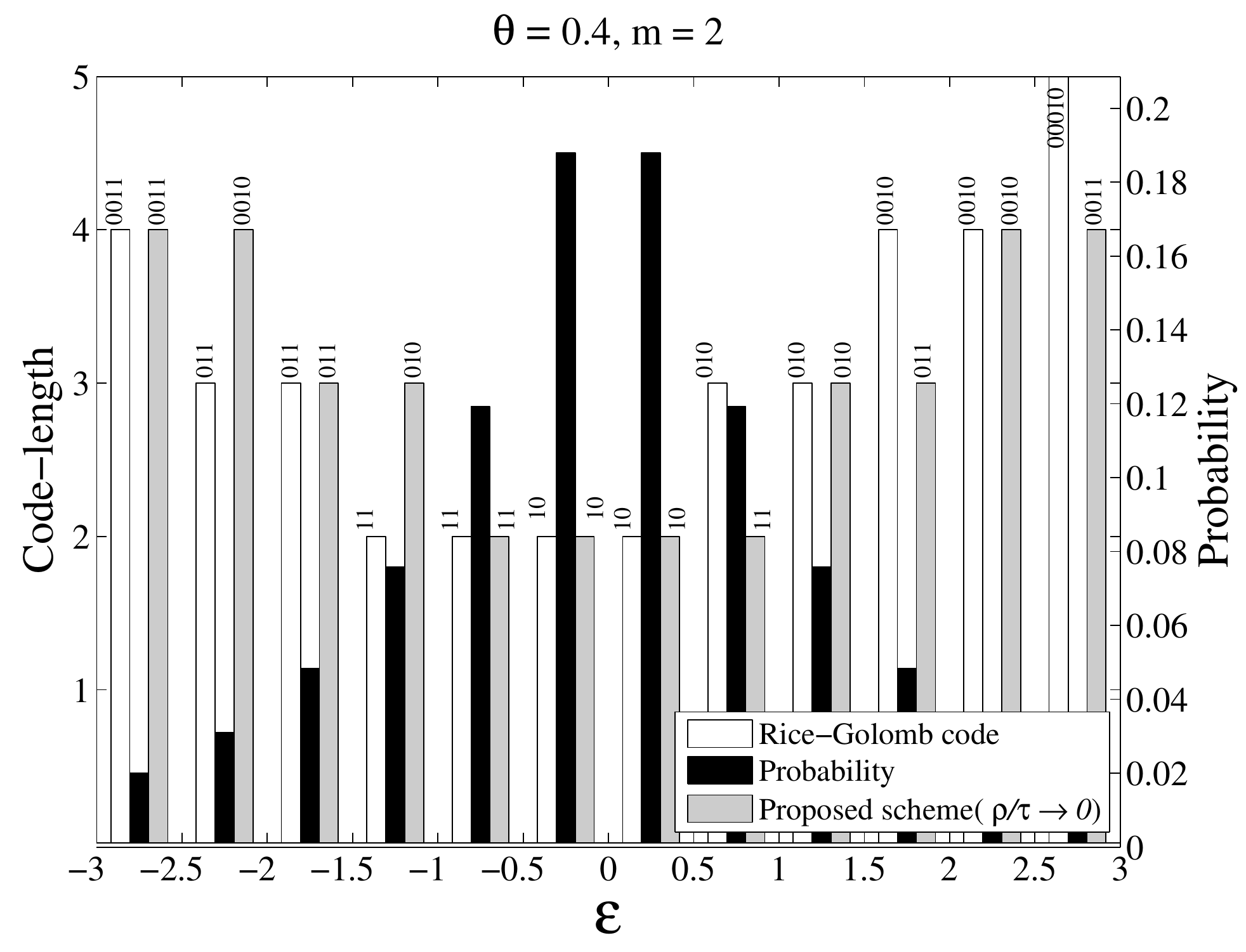}
} \\
\subfloat[][]{
\includegraphics[width=7.6cm]{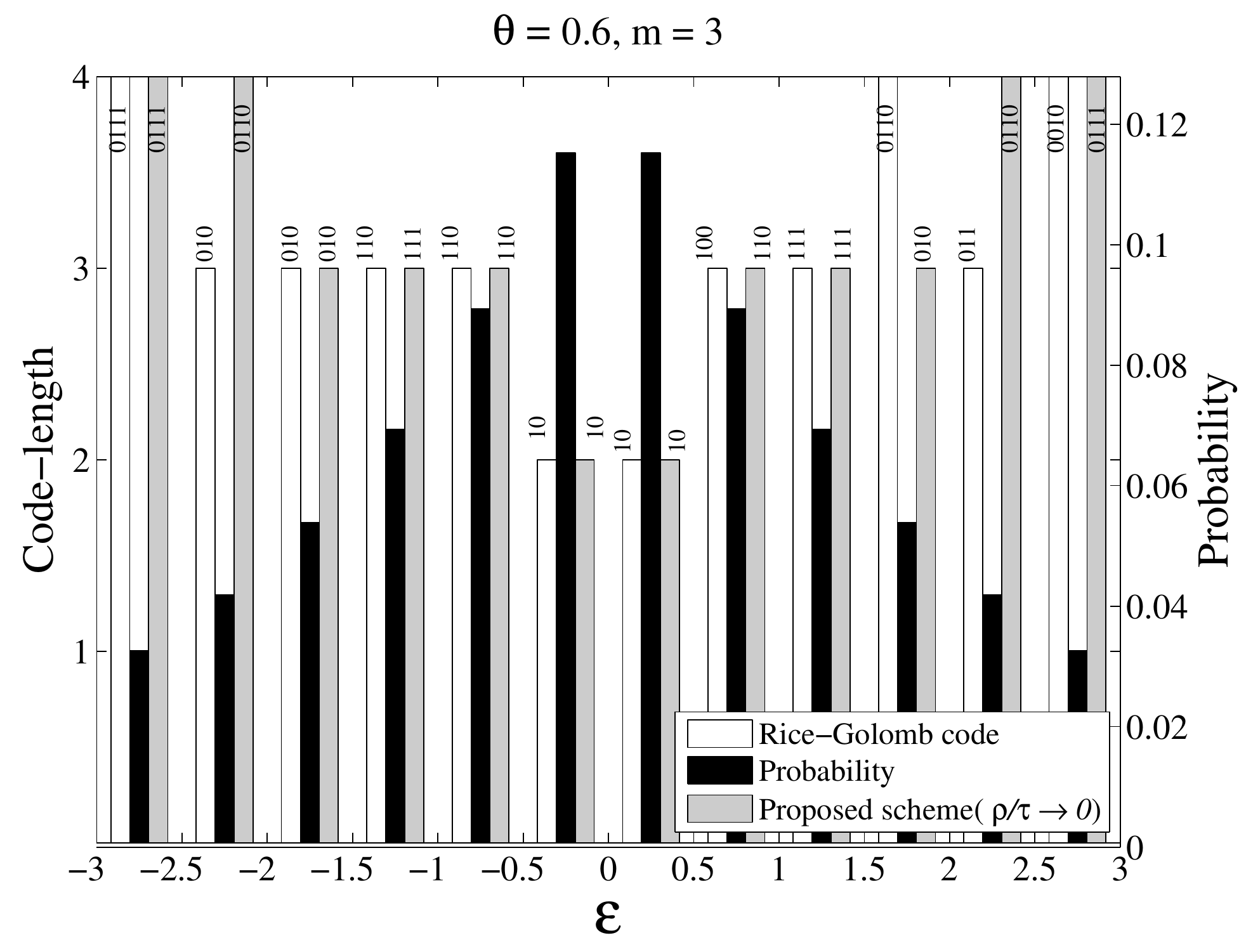}
}
\caption{Probability of intervals and the codes associated with them in both the Rice-Golomb coding and the proposed coding scheme (operating at precision $\rho/\tau \rightarrow 0$) for Laplace distributed residuals for different values of~$\theta$; (a) $\theta = 0.1$; (b) $\theta=0.4$; and (c) $\theta=0.6$.}
\label{fig:BarChart}%
\end{figure}
The code-lengths in different residual intervals for $\theta = 0.1, 0.4$, and $0.6$ are shown in Fig.~\ref{fig:BarChart} for both the Rice-Golomb scheme and the proposed scheme. Optimal values of the coding parameter for $\theta = 0.1, 0.4$, and $0.6$ are $1$, $2$, and $3$ respectively for both the schemes. While, in the proposed scheme, the code-length of a positive residual is equal to that of its negative counterpart, Rice-Golomb coding uses shorter codes for negative intervals than equally probable positive intervals. Consequently, the lengths of Rice-Golomb codes for some negative intervals are one bit shorter than the proposed codes; while the opposite holds for some of the positive intervals.  However, for each negative interval for which the Rice-Golomb code is shorter, there exists a positive interval of a higher probability for which the proposed code is shorter. For example, in Fig.~\ref{fig:BarChart}(a), the first interval on the negative side in which the Rice-Golomb code is shorter than the proposed code is $(-1.5, -1)$ having probability $0.0342$. On the other hand, the interval $(0.5, 1)$ of probability $0.1081$ is the first positive interval in which the proposed code is shorter than the Rice-Golomb code. With increasing $\theta$, however, the differences between the probabilities of positive intervals in which the proposed code is shorter and of the corresponding negative intervals in which Rice-Golomb code is shorter is reduced. Therefore, the coding gain of the proposed scheme over Rice-Golomb code also diminishes with increasing $\theta$. For example, in Fig.~\ref{fig:BarChart}(b) and Fig.~\ref{fig:BarChart}(c) the first positive intervals in which the proposed scheme has shorter code-length are $(0.5, 1)$ and $(1.5, 2)$, having probabilities $0.1162$ and $0.0524$ respectively. The corresponding negative intervals in which the Rice-Golomb scheme has shorter codes are $(-1.5, -1)$ and $(-2.5, -2)$, having probabilities $0.0735$ and $0.0406$ respectively.
\ruimte

In order to empirically evaluate the coding performance of the proposed modified Rice-Golomb scheme, $\theta$ in the range $[0.01, 0.97]$ with increments of $0.01$ and $\rho/\tau = 1, 4/5, 1/2,$ $1/4,1/5, 1/8, 1/16$, and $\rho/\tau \to 0$ were considered\footnote{ We implemented the codecs in MATLAB, whose default spacing of floating point numbers is $2.2204e-016$. Thus in our implementation $\rho/\tau=2.2204e-016$ corresponds to $\rho/\tau \to 0$.}. First we randomly generated $100,000$ uniformly distributed integers in the range $[0,127]$. For each of the integers, the associated real-valued prediction was generated so that the residuals are Laplace distributed with parameter $\theta$. Then for each $\theta$, the optimal value of the parameter $m$ for Rice-Golomb coding, which corresponds to coding at precision $\rho/\tau=1$, was determined exhaustively. However, in choosing the value of the coding parameter $m$ at other precisions, we adopted the sub-optimal strategy as discussed in the previous section. More specifically, for a given $\theta$, we used the same $m_\theta$ for $\rho/\tau = 4/5, 1/2,$ $1/4,1/5, 1/8, 1/16$, where $m_\theta$ is the optimal value of $m$ at precision $\rho/\tau \rightarrow 0$.

\begin{table}
	\centering
	\caption{The average code-lengths achieved by the proposed modified Rice-Golomb scheme at different precisions.}
		\begin{tabular}{c|cccccc} \hline
			& \multicolumn{6}{c}{Laplace scaling parameter, $\theta$} \\
			Precision & 0.1 & 0.2 & 0.3 & 0.4 & 0.5 & 0.6 \\ \hline
			$\rho/\tau = 1$	& 1.54311 &	1.90271	& 2.2863 &	2.73099	& 3.06241 &	3.46446 \\
            $\rho/\tau = 4/5$ & 1.52707	& 1.87203	& 2.25887	& 2.6775	& 3.01546	& 3.4579 \\
            $\rho/\tau = 1/2$ & 1.53743 &	1.88422 &	2.26756 &	2.67853 & 3.01862 &	3.46191 \\
            $\rho/\tau = 1/4$ &	1.47973 &	1.83218 &	2.2252 & 2.66607 &	3.00676 & 3.45299 \\
            $\rho/\tau = 1/5$ &	1.4634 & 1.81974 &	2.2114 & 2.66405 &	3.0049 & 3.45013 \\
            $\rho/\tau = 1/8$ &	1.4643 &	1.82011 &	2.21453 &	2.66147 &	3.0028 &	3.44999 \\
            $\rho/\tau = 1/16$ & 1.46171 &	1.81674 &	2.21041 &	2.66053 &	3.00278 &	3.44938 \\
            $\rho/\tau \rightarrow 0$ &	1.46074 &	1.81614 &	2.20911 &	2.66071 &	3.00176 & 3.4499 \\
            $L_{m_\theta}(\theta)$	& 1.46248 &	1.80902 &	2.21103 &	2.66667 &	3 &	3.44719 \\
 \hline
		\end{tabular}
	\label{table:code_length}
\end{table}
Table~\ref{table:code_length} presents the average code-lengths achieved with the proposed coding scheme operating at precisions $\rho/\tau = 1, 4/5, 1/2,$ $1/4,1/5, 1/8, 1/16$, and $\rho/\tau \to 0$ for some values of $\theta$. In Table~\ref{table:code_length}, we have also shown $L_{m_\theta}(\theta)$, the average code-length achievable at precision $\rho/\tau \rightarrow 0$ using the optimal value of the parameter $m$ as computed in~(\ref{eqn:asySumMRG}). At any precision, the average code-length increases with $\theta$. While for a fixed $\theta$, the average code-length decreases as the precision becomes finer, the coding gain over integer precision Rice-Golomb code diminishes with $\theta$. Note the apparent anomalies at precisions $4/5$ and $1/5$. The code-lengths at these precisions are smaller than those at precisions $1/2$ and $1/8$, respectively. When the rounded prediction $[\hat{x}]_{\rho/\tau}$ is an integer multiple of $1/2$, there are pairs of equidistant integers from the prediction; but for each pair, the integer with negative residual is ranked ahead of that with positive residual by the generalized mapping~(\ref{CommLett:Map_unique}). As both the integers are equiprobable, such a bias is inevitable, and ultimately reduces the coding efficiency. While precisions $4/5$ and $1/5$ avoids rounding the predictions to any multiple of $1/2$, this is not the case when $\tau = 2^p$. Despite having this limitation, the compression achievable at precision $1/16$ is almost the same as the compression achievable at precision $\rho/\tau \to 0$.

\begin{figure}[!tb]
\centering
\includegraphics[width=8cm]{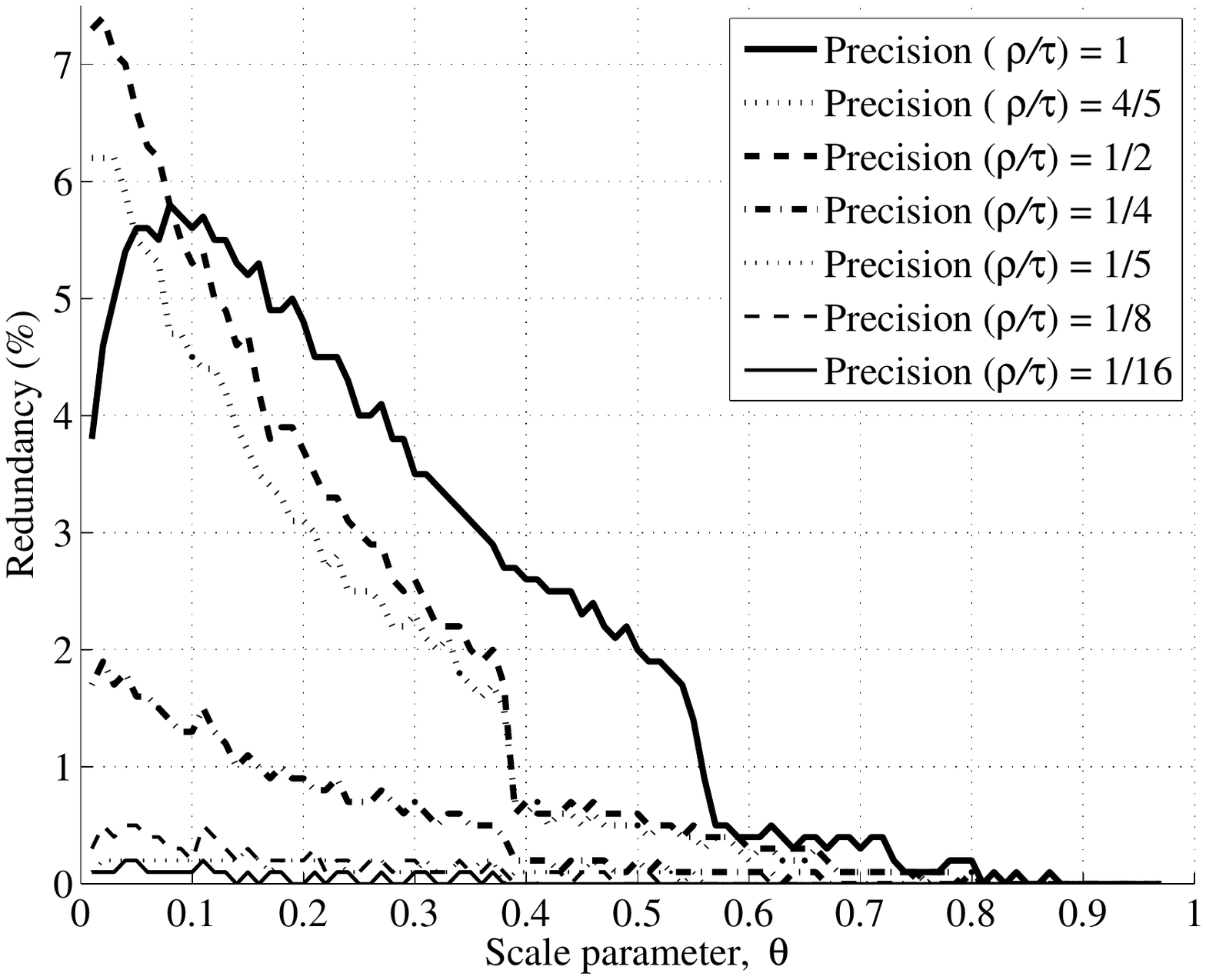}
\caption{The redundancy of the modified Rice-Golomb code at precision $\rho/\tau = 1, 4/5, 1/2,$ $1/4,1/5, 1/8, 1/16$ as compared to the average code-length achieved at precision $\rho/\tau \rightarrow 0$.}
\label{fig:ThetaTauCurve}
\end{figure}
\ruimte
In Fig.~\ref{fig:ThetaTauCurve}, we present the redundancy incurred by the proposed coding scheme operating at precisions $\rho/\tau = 1, 4/5, 1/2,$ $1/4,1/5, 1/8, 1/16$ as compared to the average code-length $L_{m_\theta}(\theta)$ achieved at precision $\rho/\tau \rightarrow 0$. It follows from the figure that, in coding Laplace distributed residuals, the redundancy incurred by the original Rice-Golomb coding scheme with unit precision is as high as 5.75\%.

\section{Conclusions}
\label{sec:con}
In this paper, the Rice-Golomb predictive coding scheme has been modified so that it can exploit the real-valued predictions more efficiently. It has been shown theoretically and demonstrated empirically that the proposed scheme achieves better compression as compared to the Rice-Golomb code when the predictions are from the real domain. Besides better compression, the proposed scheme has several advantages over the Rice-Golomb code when simplicity in analysis and implementation is of concern. Firstly, the proposed scheme allows the residuals to be modeled using the Laplace distribution instead of its discrete counterpart - the two-sided geometric distribution. Secondly, the proposed coding scheme operating at finest precision is symmetric in the sense that it assigns codewords of equal length to equally probable residual intervals. The symmetry in both the Laplace distribution and the coding scheme facilitates the analysis of the code to determine the average code-length and the optimal value of the coding parameter. Finally, the scale parameter $\theta$ of a Laplace distribution can be easily estimated online from the previous residual values. Thus we do not need to explicitly transmit the coding parameter $m$, as both the encoder and the decoder can estimate the same $\theta$ and choose the optimal $m$ given the estimated value of the $\theta$.
%%%%%%%%%%%%%%%%%%%%%%%%%%%%%%%%%%%%%%%%%%%%%%%%%%%%%%%%%%%%%%%%%%%%%%%%%%%%%%%%
%%%%%%%%%%%%%%%%%%%%%%%%%%%%%%%%%%%%%%%%%%%%%%%%%%%%%%%%%%%%%%%%%%%%%%%%%%%%%%%%

%\appendices
\appendix
%%%%%%%%%%%%%%%%%%%%%%%%%%%%%%%%%%%%%%%%%%%%%%%%%%%%%%%%%%%%%%%%%%%%%%%%%%%%%%%%
\section{Proof of the Lemmas}
\subsection{Proof of Lemma~\ref{lem:interval}}
\label{app:interval1}
Let divide the interval $[(T+1)/2 + im/2, (T+1)/2 + (i+1)m/2)$ into two sub-intervals $[(T+1)/2 + im/2, (i+1)m/2)$ and $[(i+1)m/2, (T+1)/2 + (i+1)m/2)$. Then for the first sub-interval we get,
\begin{eqnarray*}
&  & \frac{T+1}{2} + \frac{im}{2} \leq \varepsilon <  \frac{(i+1)m}{2} \\
& \Rightarrow & \frac{(T+1)}{m} + i \leq \frac{2|\varepsilon|}{m} < i+1 \\
& \Rightarrow & \left \lfloor \frac{2|\varepsilon|}{m} \right \rfloor = j = i
\end{eqnarray*}
and
\begin{eqnarray*}
&  & \frac{T+1}{2} + \frac{im}{2} \leq \varepsilon <  \frac{(i+1)m}{2} \\
& \Rightarrow & (T+1) + im \leq  \lfloor 2|\varepsilon| \rfloor < (i+1)m \\
& \Rightarrow & (T+1) \leq \lfloor 2|\varepsilon| \rfloor \textnormal{ mod } m = k < m.
\end{eqnarray*}
In this sub-interval, representation of $j$ and $k$ require $1 + i$ and $\lceil \lg m \rceil  = 1 + \lfloor \lg m \rfloor$ bits respectively. Hence,
\begin{equation}
\label{eqn:firstsub}
\ell_{m}(\varepsilon) = 2 + i + \lfloor \lg m \rfloor, \textnormal{ when } \frac{T+1}{2} + \frac{im}{2} \leq \varepsilon <  \frac{(i+1)m}{2}.
\end{equation}
For the second sub-interval we get,
\begin{eqnarray*}
&  & \frac{(i+1)m}{2} \leq \varepsilon < \frac{T+1}{2} +\frac{(i+1)m}{2} \\
& \Rightarrow & i+1 \leq \frac{2|\varepsilon|}{m} < \frac{T+1}{m} + (i+1) \\
& \Rightarrow & \left \lfloor \frac{2|\varepsilon|}{m} \right \rfloor = j = i+1
\end{eqnarray*}
and
\begin{eqnarray*}
&  & \frac{(i+1)m}{2} \leq \varepsilon < \frac{T+1}{2} +\frac{(i+1)m}{2} \\
& \Rightarrow & (i+1)m \leq 2|\varepsilon| < (T+1) + (i+1)m \\
& \Rightarrow &  0 \leq \lfloor 2|\varepsilon| \rfloor \textnormal{ mod } m = k < T+1.
\end{eqnarray*}
Thus, in the second sub-interval, representations of $j$ and $k$ require $2 + i$ and $\lfloor \lg m \rfloor$ bits respectively. Hence,
\begin{equation}
\label{eqn:secondsub}
\ell_{m}(\varepsilon) = 2 + i + \lfloor \lg m \rfloor, \textnormal{ when } \frac{(i+1)m}{2} \leq \varepsilon <  \frac{T+1}{2} + \frac{(i+1)m}{2}.
\end{equation}
Thus, when $\frac{T+1}{2} + \frac{im}{2} \leq \varepsilon <  \frac{T+1}{2} + \frac{(i+1)m}{2}$, from (\ref{eqn:firstsub}) and (\ref{eqn:secondsub}) we get,
\begin{equation*}
\ell_{m}(\varepsilon) = 2 + i + \lfloor \lg m \rfloor.
\end{equation*}
\eind
%%%%%%%%%%%%%%%%%%%%%%%%%%%%%%%%%%%%%%%%%%%%%%%%%%%%%%%%%%%%%%%%%%%%%%%%%%%%%%%%
\subsection{Proof of Lemma~\ref{lem:unique}}
\label{proof:lem:unique}
\pf
\begin{equation*}
\frac{df_n(\phi)}{d\phi} = (n+1)\phi^n + n\phi^{n-1}.
\end{equation*}
Since $\phi \in (0, 1)$ and $n>0$, $\frac{df_n(\phi)}{d\phi}>0$. Therefore, $f_n(\phi)$ is a monotonically increasing function of $\phi$. Moreover, $\lim_{\phi \to 0} f_n(\phi) = -1$ and $\lim_{\phi \to 1} f_n(\phi) = 1$ and thus $f_n(\phi)$ can have only one solution in $(0, 1)$.
\eind
\pfend
%%%%%%%%%%%%%%%%%%%%%%%%%%%%%%%%%%%%%%%%%%%%%%%%%%%%%%%%%%%%%%%%%%%%%%%%%%%%%%%%%
\subsection{Proof of Lemma~\ref{lem:intersect}}
\label{proof:lem:intersect}
\pf
There are four possible cases.
\ruimte
{\em Case 1:} $\quad m=2^{\beta_1}$ for some integer $\beta_1 \geq 0$ and $m+1 = 2 ^{\beta_2}$ for some integer $\beta_2> \beta_1$

The only possible value of $m$ for this case is $1$. Therefore,
\begin{eqnarray*}
& & L_1(\theta) = L_2(\theta) \\
& \Rightarrow & (1 - \theta^{1/2})(\theta + \theta^{1/2} - 1)=0.
\end{eqnarray*}
Since $0 < \theta < 1$, $(1-\theta^{1/2}) \neq 0$ and thus
\begin{equation}
\nonumber
\theta + \theta^{1/2} - 1=0.
\end{equation}
Substituting $\theta^{1/2} = \phi$ yields,
\begin{equation}
\nonumber
\phi^2 + \phi - 1=0.
\end{equation}
\ruimte
{\em Case 2: } $\quad m=2^{\beta_1}$ for some integer $\beta_1 \geq 0$ and $m+1 \neq 2^{\beta_2}$ for any integer $\beta_2 \geq 0$

In this case, $\lfloor \lg (m+1) \rfloor = \lg m$ and $\lceil \lg (m+1) \rceil = \lg m + 1$. Therefore,
\begin{eqnarray*}
&  & L_m(\theta) = L_{m+1}(\theta)\\
& \Rightarrow & 1 + \lg m + \frac {\theta^{\frac {m}{2}}}{1 - \theta^{\frac {m}{2}}} = 1 + \lfloor \lg (m+1) \rfloor + \frac {\theta^{\frac{2^{\lceil \lg (m+1) \rceil}-(m+1)}{2}}}{1 - \theta^{\frac{m+1}{2}}} \\
& \Rightarrow & \phi^{m+1} + \phi^{m} - 1 = 0 \textnormal{ where $\phi = \theta^{1/2}$.}
\end{eqnarray*}
\ruimte
{\em Case 3: } $\quad m \neq 2^{\beta_1}$ for any integer $\beta_1 \geq 0$ and $m+1 = 2^{\beta_2}$ for some integer $\beta_2 \geq 0$

In this case, $\lfloor \lg m \rfloor = \lg (m+1) - 1$ and $\lceil \lg m \rceil = \lg (m + 1)$. Therefore,
\begin{eqnarray*}
&  & L_m(\theta) = L_{m+1}(\theta)\\
& \Rightarrow &  1 + \lfloor \lg m \rfloor  + \frac {\theta^{\frac{2^{\lceil \lg m \rceil}-m}{2}}}{1 - \theta^{\frac{m}{2}}} = 1 + \lg (m+1) + \frac {\theta^{\frac {m+1}{2}}}{1 - \theta^{\frac {m+1}{2}}} \\
& \Rightarrow & \phi^{m+1} + \phi^{m} - 1 = 0 \textnormal{ where $\phi = \theta^{1/2}$.}
\end{eqnarray*}
\ruimte
{\em Case 4: } $\quad m \neq 2^{\beta_1}$ for any integer $\beta_1 \geq 0$ and $m+1 \neq 2^{\beta_2}$ for any integer $\beta_2 \geq 0$

In this case, $\lfloor \lg m \rfloor = \lfloor \lg (m+1) \rfloor$ and $\lceil \lg m \rceil = \lceil \lg (m + 1) \rceil$. Therefore,
\begin{eqnarray*}
&  & L_m(\theta) = L_{m+1}(\theta)\\
& \Rightarrow &  1 + \lfloor \lg m \rfloor  + \frac {\theta^{\frac{2^{\lceil \lg m \rceil}-m}{2}}}{1 - \theta^{\frac{m}{2}}} =  1 + \lfloor \lg (m+1) \rfloor + \frac {\theta^{\frac{2^{\lceil \lg (m+1) \rceil}-(m+1)}{2}}}{1 - \theta^{\frac{m+1}{2}}} \\
& \Rightarrow & \phi^{m+1} + \phi^{m} - 1 = 0 \textnormal{ where $\phi = \theta^{1/2}$.}
\end{eqnarray*}
Since according to Lemma~\ref{lem:unique}, $\phi^{m+1} + \phi^{m} - 1=0$ has a unique solution in $(0, 1)$, the lemma follows.
\eind
\pfend
%%%%%%%%%%%%%%%%%%%%%%%%%%%%%%%%%%%%%%%%%%%%%%%%%%%%%%%%%%%%%%%%%%
\subsection{Proof of Lemma~\ref{lem:mc1GEQmc}}
\label{proof:lem:mc1GEQmc}
\pf
Since $\phi_{m}$ is the root of the polynomial, $\phi^{m+1} + \phi^{m} - 1 = 0$,
\begin{equation}
\nonumber
\phi_{m}^{m+1} + \phi_{m}^{m} = 1.
\end{equation}
Since $0 < \phi_{m} < 1$,
\begin{equation}
\label{eqn:phim}
\phi_{m}^{m+2} + \phi_{m}^{m+1} < 1.
\end{equation}
However, $\phi_{m+1}$ is the root of the polynomial, $\phi^{m+2} + \phi^{m+1} - 1 = 0$ and thus,
\begin{equation}
\label{eqn:phimp1}
\phi_{m+1}^{m+2} + \phi_{m+1}^{m+1} = 1.
\end{equation}
It follows from (\ref{eqn:phim}) and (\ref{eqn:phimp1}) that $\phi_{m+1} > \phi_{m}$.
\eind
\pfend
%%%%%%%%%%%%%%%%%%%%%%%%%%%%%%%%%%%%%%%%%%%%%%%%%%%%%%%%%%%%%%%%%%
\subsection{Proof of Lemma~\ref{lem:fromabove}}
\label{proof:lem:fromabove}
\pf
For all $m \geq 1$,
\begin{equation}
\lim_{\theta \to 0} L_m(\theta) = 1 + \lfloor \lg m \rfloor.
\end{equation}
Since $\lfloor \lg (m+1) \rfloor \geq \lfloor \lg m \rfloor$, the lemma follows.
\eind
\pfend
%%%%%%%%%%%%%%%%%%%%%%%%%%%%%%%%%%%%%%%%%%%%%%%%%%%%%%%%%%%%%%%%%%
%%%%%%%%%%%%%% %%%%%%%%%%%%%%%%%%%%
\bibliographystyle{IEEEtran}
\bibliography{IEEEabrv,Thesis}
\end{document}